\documentclass[prb,twocolumn,amsmath,amssymb,showpacs,floatfix]{revtex4}

\usepackage{graphicx,bm,epsfig,setspace}
\usepackage{color}

\makeatletter
\def\graphicscale{\twocolumn@sw{0.33}{0.4}}
\makeatother

\def\spose#1{\hbox to 0pt{#1\hss}}
\def\lesssim{\mathrel{\spose{\lower 3pt\hbox{$\mathchar"218$}}
 \raise 2.0pt\hbox{$\mathchar"13C$}}}
\def\gtrsim{\mathrel{\spose{\lower 3pt\hbox{$\mathchar"218$}}
 \raise 2.0pt\hbox{$\mathchar"13E$}}}

\def\<{\langle}
\def\>{\rangle}

\newcommand*{\beq}{\begin{eqnarray}}
\newcommand*{\eeq}{\end{eqnarray}}
\newcommand*{\bea}{\begin{eqnarray}}
\newcommand*{\eea}{\end{eqnarray}}

\def\simge{\mathrel{%
       \rlap{\raise 0.511ex \hbox{$>$}}{\lower 0.511ex \hbox{$\sim$}}}}
\def\simle{\mathrel{
       \rlap{\raise 0.511ex \hbox{$<$}}{\lower 0.511ex \hbox{$\sim$}}}}

\begin{document}

\title{Accurate coarse-grained models for mixtures of colloids and \\ 
   linear polymers
   under good-solvent conditions.}

\author{Giuseppe D'Adamo}
\email{giuseppe.dadamo@sissa.it}
\affiliation{SISSA, V. Bonomea 265, I-34136 Trieste, Italy}
\author{Andrea Pelissetto}
\email{andrea.pelissetto@roma1.infn.it}
\affiliation{Dipartimento di Fisica, Sapienza Universit\`a di Roma and
INFN, Sezione di Roma I, P.le Aldo Moro 2, I-00185 Roma, Italy}
\author{Carlo Pierleoni}
\email{carlo.pierleoni@aquila.infn.it}
\affiliation{Dipartimento di Scienze Fisiche e Chimiche, Universit\`a dell'Aquila and
CNISM, UdR dell'Aquila, V. Vetoio 10, Loc.~Coppito, I-67100  L'Aquila, Italy}

\date{\today}

\begin{abstract}
A coarse-graining strategy, previously
developed for polymer solutions, is extended here 
to mixtures of linear polymers and hard-sphere colloids. In this approach
groups of monomers are mapped onto a single pseudoatom (a blob) and the effective 
blob-blob interactions are obtained by requiring the model to reproduce some 
large-scale structural properties in the zero-density limit. We show that 
an accurate parametrization of the polymer-colloid interactions is obtained 
by simply introducing pair potentials between blobs and colloids.
For the coarse-grained model in which 
polymers are modelled as four-blob chains (tetramers), the pair potentials 
are determined by means of the iterative Boltzmann inversion scheme, taking
full-monomer pair correlation functions at zero-density as targets. 
For a larger number 
$n$ of blobs, pair potentials are determined by using a simple transferability
assumption based on the polymer self-similarity. 
We validate the model by comparing its predictions with
full-monomer results for the interfacial properties of polymer solutions 
in the presence of a single colloid and for 
thermodynamic and structural properties 
in the homogeneous phase at finite polymer and colloid
density. The tetramer model is quite accurate for $q\lesssim 1$ ($q=
\hat{R}_g/R_c$, where $\hat{R}_g$ is the zero-density polymer 
radius of gyration and 
$R_c$ is the colloid radius) and reasonably good also for $q=2$. 
For $q=2$ an accurate coarse-grained description is obtained by using the 
$n=10$ blob model. We also 
compare our results with those obtained by using single-blob models with 
state-dependent potentials.
\end{abstract}


\maketitle

\section{Introduction}

Colloid dispersions are systems of great interest in several areas, 
because of their complex behavior and their many technological 
applications. Their phase behavior depends in a sensitive way on the range
of the colloid-colloid 
attraction relative to the colloid size. In the presence of very short-range
interactions only fluid-solid coexistence occurs, as is the case for the 
hard-sphere fluid. As the range of the attraction is increased, 
a fluid-fluid transition
can also occur. The addition of nonadsorbing neutral polymers 
to a colloidal dispersion provides an easy method to modify in a controlled
fashion the range of the attractive effective force  between the colloids,
hence it allows one to modify at will the phase behavior of the system. 
For dispersions of spherical 
colloids and (sufficiently long) 
nonadsorbing neutral polymers in an organic solvent,
phase behavior depends 
\cite{Poon-02,FS-02,AK-02,TRK-03,MvDE-07,FT-08,ME-09}
to a large extent only on the nature of the solvent and 
on the ratio $q \equiv \hat{R}_g/R_c$, where $\hat{R}_g$ 
is the zero-density radius of 
gyration of the polymer and $R_c$ is the radius of the colloid.
If $q$ is smaller than a critical value $q_{CEP}$, only fluid-solid coexistence 
occurs, while in the opposite case there is also 
a fluid-fluid coexistence of a 
colloid-rich, polymer-poor phase (colloid liquid) with a 
colloid-poor, polymer-rich phase (colloid gas).
Extensive theoretical and experimental work predict 
\cite{Poon-02} $q_{CEP} \approx 0.3$-0.4 for polymers under good-solvent
conditions. 

Several approaches have been used to determine the phase diagram of
colloid-polymer mixtures. On one side, several approximate thermodynamic
approaches have been used. We mention the PRISM approach,
\cite{FS-00,FS-01,RFSZ-02} density functional theory,
\cite{SDB-03,Bryk-05} and thermodynamic perturbation theory.
\cite{PVJ-03,PH-06} 
Another successful approach is free-volume
theory\cite{LPPSW-92} which has been originally developed
for mixtures of colloids and ideal polymers and later generalized
to include polymer-polymer and polymer-colloid
interactions.\cite{FT-08,ATL-02,FT-07,TSPEALF-08,LT-11}
Such an approach appears to be quite accurate\cite{FT-08,LT-11,DPP-14-GFVT}
as long as $q\lesssim 1$.
Numerical simulations of colloid-polymer systems have also been 
performed. Simulations using full-monomer representations of the 
polymers \cite{BML-03,CVPR-06,MLP-12,MIP-13} are difficult, because of the 
large number of degrees of freedom involved. As a consequence,
the simulated chains are typically relatively short. 
This implies that results are affected by significant corrections to scaling,
which must be taken into account before comparing them  with experimental data
or with results obtained in other approaches (see Ref.~\onlinecite{DPP-14-GFVT}
for a discussion).  To avoid these difficulties, coarse-grained (CG)
approaches have been developed. In these models, short-scale 
degrees of freedom of the polymer subsystem are integrated out, 
providing a simpler representation of the polymers 
that still allows one to obtain 
accurate results  for the thermodynamics of the system and for large-scale
(i.e., on length scales comparable to the
polymer size) structural properties in some range of densities and of 
polymer-to-colloid size ratios. Beside the obvious 
advantage from the computational side, CG models are also very 
convenient since they provide directly 
thermodynamics and structural properties  in the universal,
scaling limit without requiring additional extrapolations. 
For this purpose it is enough to determine the target properties
on which the CG model is built in the scaling limit. 

Several CG models have been introduced for
polymer solutions in different concentration regimes.  
\cite{Likos-01,LBHM-00,BLHM-01,PCH-07,Pelissetto-09,DPP-12-Soft,DPP-12-JCP,%
VBK-10,CG-10,KPG-10,CS-10,ZMSDK-14}
In the simplest approaches (single-blob representations), one maps 
polymer chains onto point particles
interacting by means of spherically symmetric potentials.
\cite{Likos-01,LBHM-00,BLHM-01} In the good-solvent regime,
models with density-independent potentials, i.e., obtained at
zero polymer density, reproduce the thermodynamic
behavior of polymer solutions up to polymer volume fractions
$\phi_p$ of order 1 ($\phi_p = 4\pi \hat{R}^3_g \rho_p/V$,
where $\rho_p$ is the polymer number density and $V$ is the volume), 
i.e., as long as
polymer-polymer overlaps are rare, so that the neglected
many-body interactions\cite{BLH-01} are not relevant. 
These simple models have been extended to include polymer-colloid interactions.
\cite{LBMH-02,LBMH-02-bis,BLH-02,FBD-08,Dzubiella-etal-01,DLL-02,VJDL-05}
For colloid-polymer systems, these approaches are generically expected to
be predictive only in the colloid regime $q\lesssim 1$. Indeed, for $q\gtrsim
1$, corresponding to $\hat{R}_g\gtrsim R_c$, the polymer can wrap around
the colloid, making the monoatomic (single-blob) approximation for the 
polymers unrealistic.

Single-blob models with potentials dependent
on the polymer density have also been considered.\cite{LBHM-00,BLHM-01,BL-02} 
Such models, by definition,
reproduce some large-scale properties of the system for all values of 
$\phi_p$ in the limit of zero colloidal density. In this approach, however,
there are several sources of ambiguity. First, potentials do not only depend
on the state point one considers, but also on the chosen
ensemble,\cite{DPP-13-state-dep} i.e., on the thermodynamic variable
(for instance, the density or the chemical potential) chosen to specify
the thermodynamic state point. Moreover, 
also the determination of the thermodynamic properties
is ambiguous, since different approaches, which are equivalent for 
systems with state-independent interactions, provide different results 
for the same quantity.\cite{SST-02,Louis-02,DPP-13-state-dep} 
Since potentials are not allowed to depend on the colloidal density $\rho_c$,
for finite $\rho_c$ this approach only provides an approximation to the 
correct behavior, with the same limitations of the simpler 
zero-density single-blob approach. In particular,
also this class of single-blob models is expected to be predictive
only for $q\lesssim 1$.

These limitations can be
overcome by switching to a model at a lower level of coarse graining:
A long polymer chain is mapped onto a short chain of $n$ soft monomers
(blobs).\cite{PCH-07,DPP-12-Soft,DPP-12-JCP} For purely polymeric systems, 
these CG models predict the correct thermodynamic and large-scale
structural properties of the solution as long as blob overlaps are rare.
\cite{PCH-07,DPP-12-JCP}
In the semidilute regime this condition is satisfied for
$\hat{r}_g\ll \xi$, where $\hat{r}_g$ is the zero-density 
blob radius of gyration and $\xi$ is the de Gennes correlation length,
which is the only relevant length scale in the semidilute regime.
Since\cite{DPP-12-Soft} $\hat{r}_g/\hat{R}_g \sim n^{-\nu}$ and 
\cite{deGennes-79}
$\xi/\hat{R}_g \sim \phi_p^{-\gamma}$, where $\gamma = \nu/(3\nu-1)$ 
and $\nu$ is the Flory exponent ($\nu = 0.587597(7)\approx 3/5$, see
Ref.~\onlinecite{Clisby-10}), the condition $\hat{r}_g\ll \xi$ 
is equivalent to
$\phi_p \lesssim n^{3\nu - 1} \approx n^{0.76}$ under good-solvent 
conditions.
Therefore, by increasing the number $n$ of blobs, the CG model
becomes predictive in a larger density interval. Analogously, in the 
presence of colloids, the pairwise approximation in which many-body 
interactions are neglected only holds if
$\hat{r}_g/R_c \lesssim 1$, i.e., for $q\lesssim n^\nu$. This condition 
guarantees that the average distance 
between two colloids is  larger than $\hat{r}_g$ and that, therefore, 
only pair interactions are relevant.  Again, by increasing 
the number $n$ of blobs we can extend the validity of CG models to 
larger values of $q$.

Recently, we have introduced a procedure to set up a hierarchy of CG models for
linear polymer chains under good-solvent conditions which simultaneously 
reproduce quite accurately
structure and thermodynamics of polymer solutions deep into the semidilute
regime.\cite{DPP-12-Soft,DPP-12-JCP}
We followed the structure-based route
\cite{MullerPlathe-02,Reith:2003p2128,PK-09,KVMP-12,BAGLRRV-13} 
(an alternative, conceptually different approach is the force-matching
route, discussed, e.g., in Refs.~\onlinecite{force-matching1,YBGCG}),
in which 
the CG model is set up in such a
way to reproduce full-monomer correlation functions of a set of chosen
structural collective variables, which are
determined at zero polymer density. Since simulations of a few isolated
polymer chains are relatively easy,
we were able to obtain the target correlation functions 
for polymer chains of several lengths with high numerical precision.
Therefore, we could perform a reliable extrapolation to obtain 
target properties in the scaling, universal limit. This guarantees that 
the CG model gives thermodynamical and structural predictions in the 
scaling limit, which can directly be compared with experiments on
high-molecular-weight polymers.

In the minimal model each linear chain is represented by a short
polyatomic molecule with four sites (tetramer). The tetramer
potentials are set up at zero density---this allows us to avoid
the inconsistencies\cite{SST-02,Louis-02,DPP-13-state-dep} that occur when
using state-dependent potentials---by matching the single-chain
intramolecular structure and the center-of-mass pair correlation function
between two identical chains. This minimal representation has
been shown to provide accurate results for the underlying solutions
up to $\phi_p\simeq2$.\cite{DPP-12-Soft} Higher-resolution models
with $n > 4$ are obtained by
using a simple transferability approach, which allows one to obtain
the interaction potentials for $n$ blob systems starting from those
computed for the tetramer. 
This transferability
approximation, extensively discussed in Sec.~V.A of 
Ref.~\onlinecite{DPP-13-thermal}, 
was shown to be quite accurate\cite{DPP-12-JCP} and allowed
us to obtain precise thermodynamic and (large-scale) structural
results for $\phi_p\gg 1$. For instance, the multiblob CG models predict
the isothermal compressibility with an error of less then 10\% up
to $\phi_p\approx 2$ for $n=4$, $\phi_p\approx 4.5$ for $n=10$ and 
up to $\phi_p\approx 10$ for $n=30$.

In this paper we extend the multiblob approach to colloid-polymer mixtures.
Colloids are modelled as hard spheres of radius $R_c$, while a multiblob
model is used for polymers.
First, we consider the case in which each polymer is represented 
by a four-blob (tetramer) CG molecule.
The resulting CG model, in which polymer-colloid
interactions are simply approximated by pair potentials between the blobs 
and the hard sphere, works quite well up to $q\approx 2$ in the homogeneous
phase and represents a significant improvement with respect to the 
single-blob case. Then, we extend the model to $n > 4$ by using a simple
transferability argument, analogous to that presented in
Ref.~\onlinecite{DPP-12-JCP}.
To validate the model, we compare the numerical data 
obtained by using the CG models 
with full-monomer simulation results  for $q=0.5$, 1, and 2. 
Beside being of relevance for the test of the CG model, these simulations also 
provide new results for the intermolecular and intramolecular structure 
in a colloidal dispersion.

The paper is organized as follows. In Sec.~\ref{sec2} we define 
our basic CG model in which each polymer is represented
by a four-blob (tetramer) CG molecule. In Sections \ref{sec3}, \ref{sec4},
and \ref{sec5} we determine the accuracy with which the tetramer model
reproduces the behavior of the polymer-colloid mixture. First, we consider the 
third virial coefficients, that allow us to estimate how large 
the neglected three-body effects are.\cite{DPP-12-compressible,AW-14} Then,
we consider the behavior of a single colloid in a bath of polymers. 
In Sec.~\ref{sec5} we present full-monomer
results in the homogeneous phase for $q=0.5$, $q=1$ and $q=2$. 
We determine several thermodynamic quantities, which are then compared with 
the corresponding tetramer results. In Sec.~\ref{sec6} we consider 
the transferability in the number of blobs, determining a decamer model,
which is validated by using the full-monomer results derived in the preceding 
sections. In Sec.~\ref{sec7} we discuss single-blob models with 
state-dependent interactions, generalizing that proposed in  
Ref.~\onlinecite{LBHM-00,BLHM-01,BLH-01,BLH-02}. Finally, in Sec.~\ref{sec8}
we draw our conclusions. In App.~\ref{Appendix-KB} we collect some 
formulae that are useful to determine some thermodynamic quantities 
from Monte Carlo estimates of the partial structure factors. 
In App.~\ref{App.B} we discuss the polymer and blob size in the 
homogeneous phase as a function of $\phi_c$ and $\phi_p$.
In App.~\ref{App.C} we explain how to determine state-dependent single-blob
potentials in the grand-canonical ensemble by using integral-equation methods.
Details are collected in the supplementary material.\cite{suppl}
We define the polymer model we use to compute full-monomer properties,
we give interpolations of the colloid-blob potentials determined for 
the tetramer model (for the pure polymer system, see the supplementary material
of Ref.~\onlinecite{DPP-13-thermal}), and provide extensive tables of 
thermodynamic data in the homogeneous phase. 

\section{The coarse-grained model} \label{sec2}

\subsection{Definitions}

In the multiblob approach one starts from a {\em coarse-grained representation}
(CGR) of the underlying full-monomer model, which is obtained by
mapping a chain of $L$ monomers onto a chain of $n$ blobs, each of them
located at the center of mass of a subchain of $m=L/n$ monomers.
If the monomer positions are given by
$\{ {\bf r}_1,\ldots, {\bf r}_L\}$, one first defines the
blob positions ${\bf s}_1,\ldots, {\bf s}_n$ as the
centers of mass of the subchains of length $m$, i.e.
\begin{equation}
 {\bf s}_i = {1\over m} \sum_{\alpha=m(i-1)+1}^{mi}  {\bf r}_\alpha.
\label{si-def}
\end{equation}
For the new CG chain $\{{\bf s}_1,\ldots, {\bf s}_n\}$
one defines several intramolecular and intermolecular correlation functions,
which are then used as target properties for the definition of the CG model.

The CG model consists of polyatomic molecules of $n$ atoms located in
$\{{\bf t}_1,\ldots,{\bf t}_n\}$. All length scales are expressed in terms
of the full-monomer zero-density radius of gyration $\hat{R}_g$, 
hence all potentials and
distribution functions depend on the adimensional combinations
${\bf b} = {\bf t}/\hat{R}_g$. As discussed in
Ref.~\onlinecite{DPP-12-Soft}, it is not feasible to reproduce exactly the 
structure of the full-monomer model, even at the CGR level, since this
would require the introduction of complex many-body interactions. 
However, a judicious parametrization of the interactions 
in terms of pair potentials, each of them
depending on a single scalar variable, works quite well.\cite{DPP-12-Soft}
In this paper we apply the same strategy to polymer-colloid mixtures,
determining the appropriate effective intermolecular interactions between
polymer blobs and colloids.
In the spirit of the multiblob approach,
we neglect interactions among three or more molecules and consider only
the interaction 
between a polymer and a colloid. In general, it depends on $3(n-1)$
scalar variables, parametrizing the relative position of the blobs and of the
colloid. The exact determination of this many-body potential is, of course,
unfeasible in practice.
Therefore, 
we make a pair-potential approximation. The polymer-colloid interaction
is completely specified by the blob-colloid pair potentials 
$V_{cp,i}(b,q)$, where $i$ is the blob index along the chain,
$b = |{\bf r}_c - {\bf t}_i|/\hat{R}_g$, and ${\bf r}_c$, ${\bf t}_i$ are 
the colloid and blob positions, respectively. Note that the potentials depend 
on $q$, which is explicitly reported in the notation.

As in Ref.~\onlinecite{DPP-12-Soft}, we begin by considering the 
tetramer case $n = 4$. For this value of $n$ there are two independent
potentials $V_{cp,1}(b,q)=V_{cp,4}(b,q)$ and 
$V_{cp,2}(b,q)=V_{cp,3}(b,q)$. They are determined by requiring the CG
model to reproduce the distribution functions $g_{cp,i}(b,q)$ 
between the center of mass of blob $i$ and the colloid,
where $r = |{\bf r}_c - {\bf s}_i|$ and $b = r/\hat{R}_g$.
For $n>4$ a direct computation of the potentials 
is unfeasible, hence we will use a simple transferability approach as in 
Ref.~\onlinecite{DPP-12-JCP}.

\begin{figure}[tb!]
\begin{center}
\begin{tabular}{c}
\includegraphics[width=7cm]{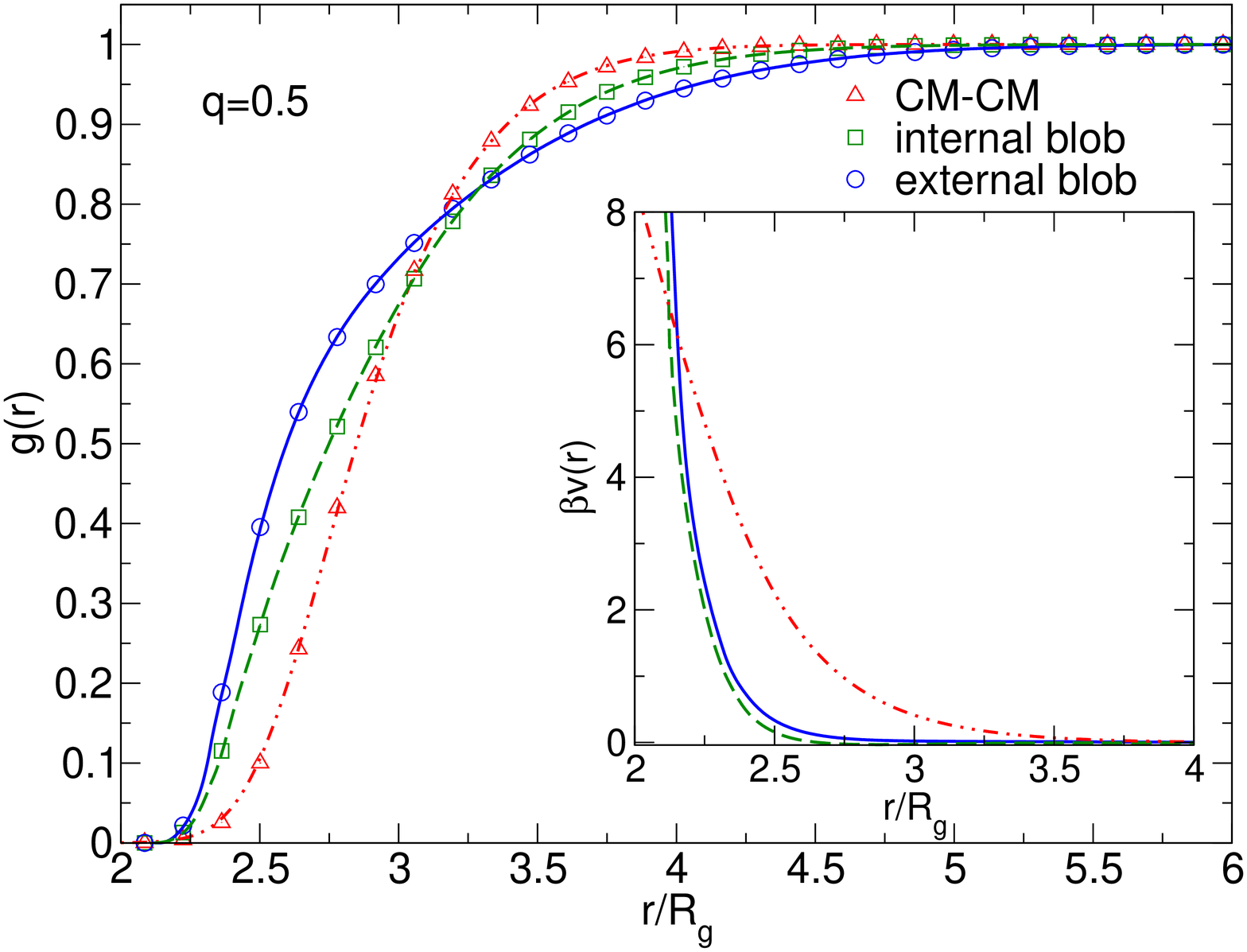} \\
\includegraphics[width=7cm]{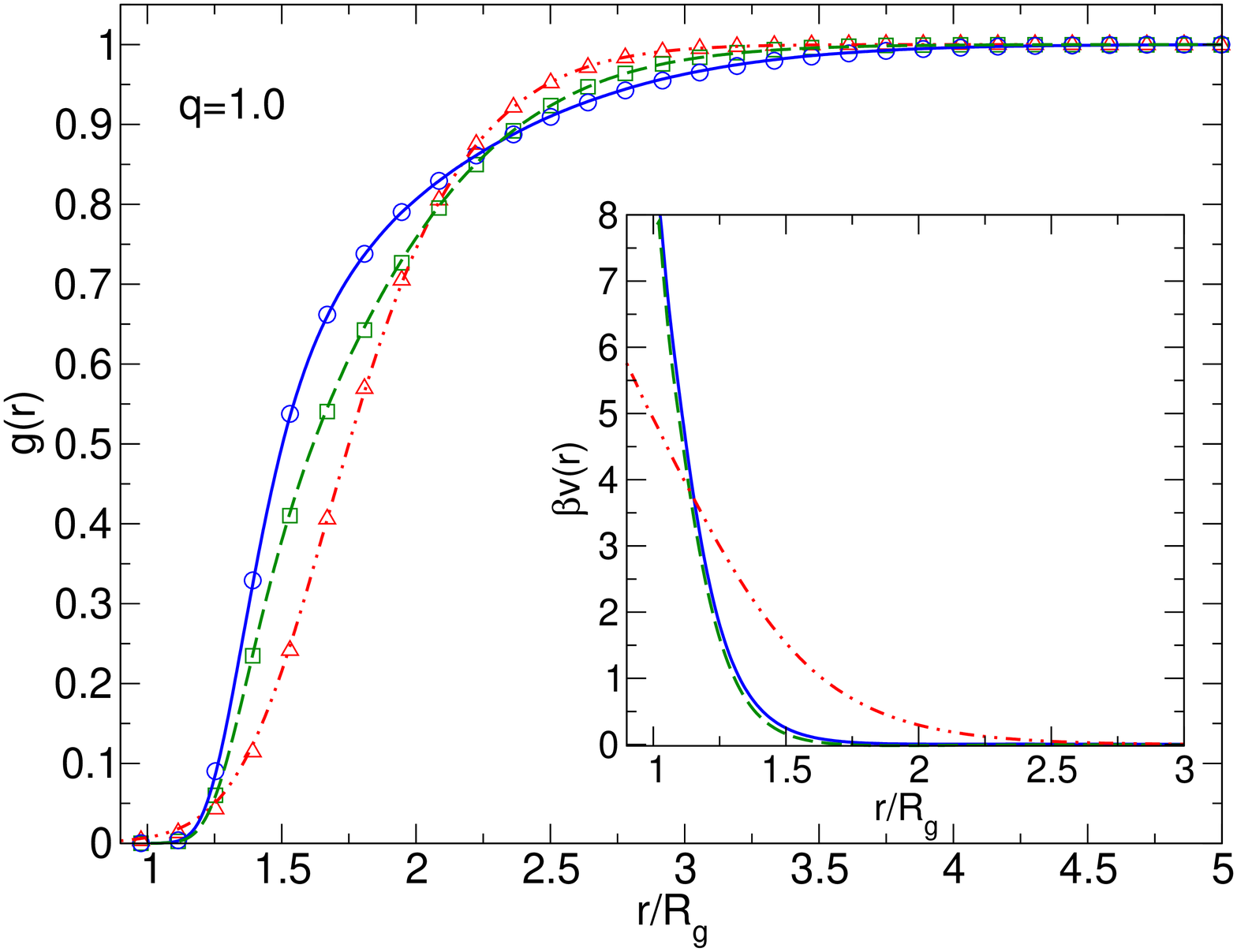} \\
\includegraphics[width=7cm]{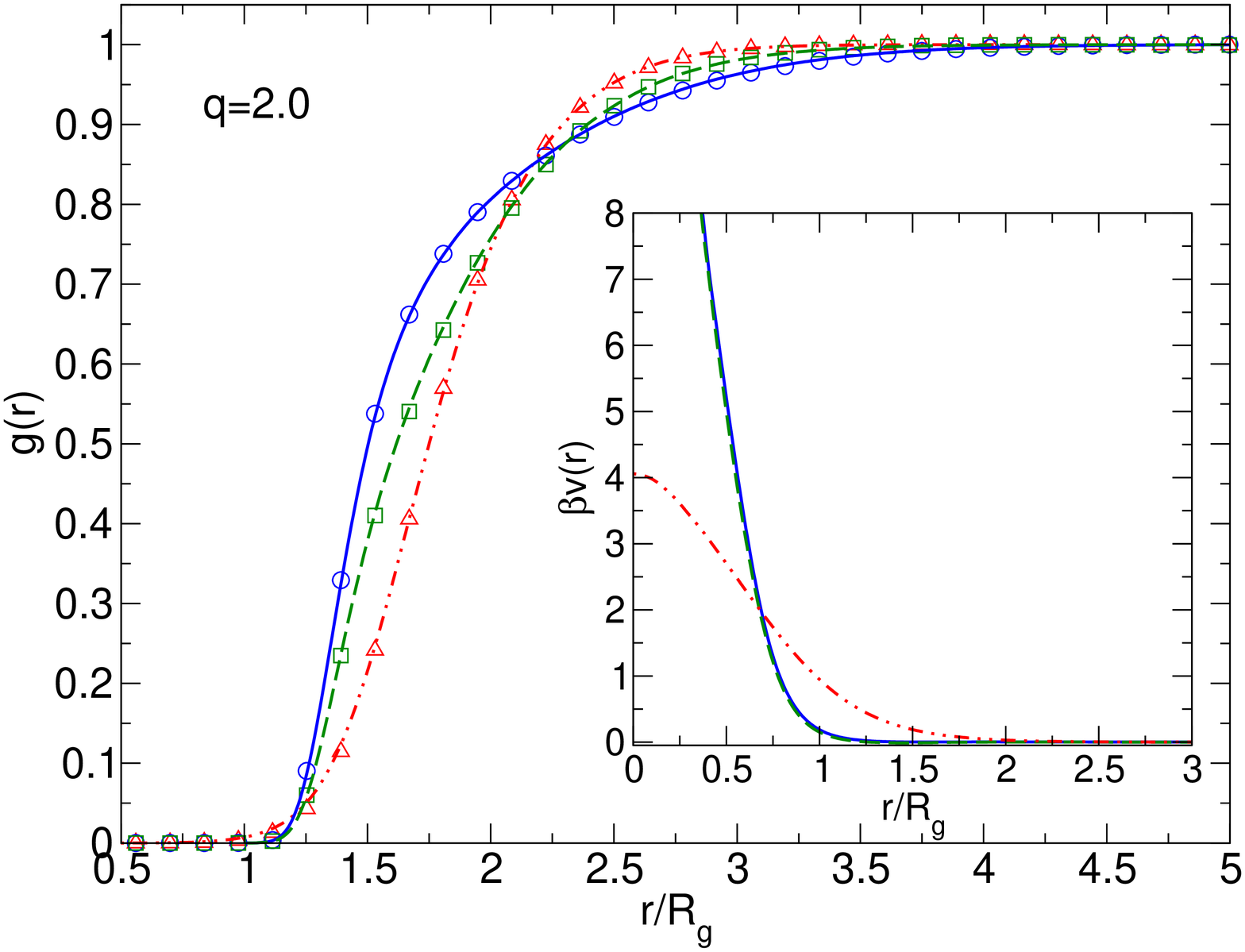} \\
\end{tabular}
\end{center}
\caption{Pair distribution functions between the centers of mass of 
the blobs and the colloid  (solid lines and circles 
refer to the external blobs, dashed lines and squares to the 
internal blobs) and between the center of mass of the polymer and the colloid 
(dash-dot-dot line and triangles). Symbols refer to the CG estimates,
while the lines are the corresponding full-monomer results.
In the inset the corresponding potentials are shown with the same line
conventions as for the distribution functions. Results for $q=0.5$ (top),
1 (middle), and 2 (bottom).
}
\label{fig:potenziali-gr}
\end{figure}

\subsection{Tetramer polymer model}

In order to determine full-monomer properties,
we consider the three-dimensional lattice Domb-Joyce
model \cite{DJ-72} as in our previous work (see the supplementary
material\cite{suppl} for the precise definition). 
In the absence of colloids, there is a significant
advantage in using Domb-Joyce chains instead of other models. For 
a generic polymer model, the leading 
scaling corrections decay slowly, as $L^{-\Delta}$ ($\Delta = 0.528(12)$,
Ref.~\onlinecite{Clisby-10}), where $L$ is the number of monomers of 
each chain. Therefore, the
universal, large--degree-of-polymerization limit is only observed for
quite large values of $L$. 
On the other hand, if the repulsion parameter that appears in the 
Domb-Joyce Hamiltonian is chosen appropriately, the scaling 
corrections decay faster,
approximately as $1/L$.\cite{BN-97,CMP-06}
As a consequence, scaling results are obtained
by using significantly shorter chains (for zero-density quantities,
simulations of chains with 600 monomers give results 
that are essentially asymptotic). Unfortunately, in the presence of
a repulsive surface, new renormalization-group operators arise, which are
associated with the surface.\cite{DDE-83}
The leading one gives rise to corrections that scale as $L^{-\nu}$,
\cite{DDE-83} where $\nu$ is the Flory exponent (an explicit test
of this prediction is presented in the supplementary material 
of Ref.~\onlinecite{DPP-13-depletion}), hence it
spoils somewhat the nice scaling behavior observed in the absence of colloids.
These corrections are not negligible, even for chain lengths of the order of 
$10^3$. Therefore, finite-length polymer-colloid results must be extrapolated
to obtain scaling results.

In practice, we work as follows. We consider Domb-Joyce chains of length
$L = 240$, $L = 600$, and $L = 2400$. For each value of $q$ and $L$ 
we determine the blob-colloid distribution functions $g_{cp,i}(r,L,q)$ 
by Monte Carlo simulations.
In the scaling limit $L\to \infty$, these quantities converge to universal
functions, once distances are expressed in units of the size of the polymer,
i.e., in terms of the adimensional ratio $b = r/\hat{R}_g$.
Therefore, for each $L$ we interpolate 
$g_{cp,i}(b, L, q)$ by means of a cubic spline.
Then, the data for the three values of $L$ are extrapolated by assuming
$g_{cp,i}(b, L, q) = a(b,q) + c(b,q) L^{-\nu} + d(b,q)/L$. Solving the 
simple linear system, we obtain $a(b,q)$ and set 
$g_{cp,i}(b,q) = a(b,q)$. This quantity is then used as target
distribution function.

Once the target functions are known, 
the potentials are obtained
by applying the Iterative Boltzmann Inversion (IBI) scheme.
\cite{Schommers:1983p2118,MullerPlathe-02,Reith:2003p2128}
Less than ten iterations are needed to reproduce the target 
quantities quite precisely.
In Fig.~\ref{fig:potenziali-gr} 
we show the blob-colloid pair distribution functions and the 
corresponding effective potentials for the 
tetramer for  three values of the size ratio, $q=0.5,1.0,2.0$. The 
potentials are short ranged and become very small approximately for 
$r/\hat{R}_g \approx 1/q + 0.5$. This is consistent with the idea that 
the typical range is of the order of $R_c+\hat{r}_g$, since 
$\hat{r}_g \approx 0.45 \hat{R}_g$ for a tetramer.\cite{DPP-12-Soft} Moreover,
they increase  steeply  as $b$ approaches the contact point 
$R_c/\hat{R}_g=1/q$. For $q=0.5$ and 1, colloids and blobs cannot 
approach each other by less than $R_c$ and indeed, the potentials 
apparently diverge when $b\to 1/q$. 
On the other hand, for $\hat{r}_g\gtrsim R_c$
the blobs can wrap around the colloids, hence there is a finite probability
that the distance between the blob and the colloid is less than $R_c$. 
This occurs for $q=2$ (we have\cite{DPP-12-Soft}
$\hat{r}_g \approx 0.95 R_c$ in this case). For this value of $q$,
$\beta V_{cp,i}(1/q,2) \approx 5$.

The difference between the potentials associated with the internal and 
external blobs---the internal-blob potential is less repulsive than
the external one---is
small but not negligible, especially
for $q = 0.5$, and decreases with $q$. 
An attractive tail  of the 
order of $10^{-2}k_B T$ is also observed but we believe it is a numerical 
artifact of the IBI procedure, as reported in other contexts.
\cite{MullerPlathe-02}

\section{Comparison of full-monomer and tetramer results at zero-density}
\label{sec3}

\begin{table*}[tbp!]
\caption{Virial combinations for full-monomer (FM) systems (scaling-limit
values from Ref.~\onlinecite{DPP-13-depletion}), for the single-blob
model ($n=1$), for the tetramer ($n=4$), and 
for the decamer ($n=10$, method c)). 
For the third-virial 
combinations, we also report the simple-liquid contribution $A_{3,\#}^I$ and
the flexibility contribution $A_{3,\#}^{\rm fl}$ (see
Ref.~\onlinecite{DPP-13-depletion}, App. A, for the definitions): 
$A_{3,\#} = A_{3,\#}^I + A_{3,\#}^{\rm fl}$.}
\label{tab:virial-tetramer}
\begin{center}
\begin{tabular}{ccccccccc}
\hline
\hline
$q$ & $n$ & $A_{2,cp}$ & $A^I_{3,cpp}$ & $A^{\rm fl}_{3,cpp}$ & $A_{3,cpp}$  & 
                   $A^I_{3,ccp}$ & $A^{\rm fl}_{3,ccp}$ & $A_{3,ccp}$ \\
\hline
0.5 & FM & 107.4(3) & 748(4) & $-$22(2)&726(5)& 8759(45) & $-$130(6)&8630(45) \\
    & 1	 & 107.253(6)&704.7(4)& 0  & 704.7(4) & 8621(2)  & 0      & 8621(2) \\
    & 4  & 107.49(3)&757.7(6)&$-$16(1)& 742(2)&8771(14)&$-$120(6)&8651(13)\\
\hline
1   & FM & 27.54(6) & 140.0(8)&$-$6.8(2)& 133.3(9) &371(2)&$-$12.2(5) &360(2) \\
    & 1  & 27.289(2) & 119.53(8)& 0   & 119.53(8) & 340.6(2) & 0  & 340.6(2) \\
    & 4	 & 27.70(1)  & 140.0(5) &$-$4.8(4)& 135.3(6)&370(2)&$-$10.6(6)&359(2) \\
  & $10$& 27.592(7)&143.6(2)&$-$6.0(2)&137.6(3)&374.9(4)&$-$11.8(4)&363.1(4)\\
\hline
2   & FM & 8.65(5) & 28.1(2) & $-$2.0(1) & 26.1(2)&17.80(2)&$-$1.03(5)&16.8(1)\\
    & 1	 & 8.6049(9) & 13.25(3)& 0 &  13.25(3) & 20.36(2) & 0 & 20.36(2) \\
    & 4	 & 8.679(3)  & 27.3(1) & $-$1.3(1) &26.0(2)&17.4(2)&$-$0.81(6)&16.6(2)\\
  & $10$&8.687(5)  & 28.8(1) & $-$1.8(2)&27.0(3)&18.0(1)&$-$0.8(1)&17.0(2)\\
\hline
\hline
\end{tabular}
\end{center}
\end{table*}

We wish now to discuss the behavior of the CG model in the thermodynamic
regime 
in which both the colloidal and the polymer densities $\rho_p=N_p/V$
and $\rho_c=N_c/V$ are small. In this limit the thermodynamic 
properties of interest can be expressed as a series expansion in the 
concentration variables. The coefficients of these expansions can be 
typically related to the virial coefficients parametrizing the concentration 
dependence of the (osmotic) pressure $P$. At third order in the 
densities we have 
\begin{eqnarray}
&& \beta P \approx \rho_c+\rho_p+B_{2,cc}\rho_c^2+B_{2,cp}\rho_c\rho_p+
B_{2,pp}\rho_p^2 \nonumber \\
&& \\
&& \quad 
  +B_{3,ccc} \rho_c^3+B_{3,ccp} \rho_c^2 \rho_p+B_{3,cpp} \rho_c \rho_p^2+
    B_{3,ppp} \rho_p^3 \nonumber 
\end{eqnarray}
Although virial coefficients are model dependent, their adimensional 
combinations $A_{2,\#} = B_{2,\#}\hat{R}_g^{-3}$ and 
$A_{3,\#} = B_{3,\#}\hat{R}_g^{-6}$ are universal. Hence, it makes 
sense to compare full-monomer predictions with CG results. Since the
CG model has been defined by matching the blob-colloid distribution 
functions, the compressibility rule\cite{HMD-06} implies that 
$A_{2,cp}$ should be the same in the full-monomer and in the CG model.
Hence, the comparison of $A_{2,cp}$ allows us to verify the 
accuracy of the inversion procedure. 
Results
 for $A_{2,cp}$ are reported in Table~\ref{tab:virial-tetramer}.
\cite{footnote-viriali} 
In all cases $A_{2,cp}$ is close to the full-monomer estimate, 
confirming the validity of the CG potentials. The comparison of the 
estimates of the third-virial combinations is much more interesting, 
since it allows us to estimate how effective the CG model is in modelling
three-body interactions, \cite{DPP-12-compressible,AW-14}
which are relevant in 
the concentration regimes in which multiple overlaps between polymers and 
colloids cannot be neglected.

In Table~\ref{tab:virial-tetramer} we collect results for the 
third-virial combinations
obtained from full-monomer simulations\cite{DPP-13-depletion} and for 
the CG model at various levels of resolution. For future convenience,
we also present results for the CG model with $n=10$ blob, the decamer, a CG 
model that will be discussed in Sec.~\ref{sec6} (the same comment applies to 
the Tables and figures that will be presented below).
As it was shown in Ref.~\onlinecite{DPP-13-depletion}, App. A,
for polyatomic molecules these quantities are 
the sum of two contributions. One contribution, that we denote with 
$A^I_{3,\#}$ is the usual term that gives the third virial coefficient
in simple liquids of monatomic molecules (diagramatically, it is associated
with the triangle diagram\cite{HMD-06}). The second contribution, denoted with 
$A^{\rm fl}_{3,\#}$, is a flexibility contribution that takes into
account the conformational properties of the polymers. It represents
a small, but not negligible correction to $A^I_{3,\#}$, which becomes
more important as $q$ increases. Both quantities are universal, hence 
a separate comparison is meaningful.

As already observed in the purely polymeric case,\cite{DPP-12-Soft}
the tetramer model reproduces well the third virial combinations,
indicating that three-body interactions are correctly taken into account,
at least up to $q\approx 2$.
The relative difference between the tetramer and full-monomer estimates of 
$A_{3,ppc}$ 
is $1\%$ and $2\%$ for $q=1$ and $q=2$, respectively.  Similar 
observations hold for $A_{3,ccp}$ that is accurately 
reproduced by the CG model: differences are at most of 
$0.3\%$ and $1.5\%$ for $q=1$ and $q=2$, respectively. Note that 
the three-body effects involving two colloids and one polymer are 
better reproduced than those involving two polymers and one colloid.
The data of Table~\ref{tab:virial-tetramer} also show that
the tetramer represents a significant improvement 
with respect to the single-blob model. First, the latter is unable to reproduce 
the flexibility correction. Second, deviations from the 
correct, full-monomer results are quite significant:
$A_{3,ccp}$ and $A_{3,cpp}$ are underestimated by 10\% and 5\% for $q=1$,
respectively. For $q=2$ deviations are significantly larger: $A_{3,cpp}$
differs by a factor of two from the full-monomer result.

As an additional check of the validity of the procedure we have compared
the distribution function $g_{cp,CM}(b,q)$ between the colloid and 
the polymer center of mass computed in the full-monomer and in the 
CG model. Since we used the blob-colloid distribution functions as targets
for the inversion procedure, this is a nontrivial check that 
allows us to verify how good the pair-potential approximation is for 
the intermolecular interactions. 
The results are shown in Fig.~\ref{fig:potenziali-gr}. 
In all cases, we observe a very good agreement, confirming the accuracy of the
procedure.

\section{Full-monomer and tetramer results for a  
pure polymer solution in the presence of a spherical solute}
\label{sec4}

\begin{table}[tbp!]
\caption{Depletion thickness $\delta_s(\phi_p)/R_c$ as a function of the 
polymer volume fraction $\phi_p$. 
Full-monomer (FM), single-blob ($n=1$), 
tetramer ($n=4$), and decamer ($n=10$) results.
}
\label{tab:deplection-vs-phip}
\begin{center}
\begin{tabular}{cccccc}
\hline\hline
\squeezetable
$q$ & $\phi_p$ & FM & $n=1$ & $n=4$ & $n=10$\\
\hline
0.5	   &   0.0 &	0.474(1)  &  0.47371(2) &  0.4745(1)& \\
	   &   0.4 &	0.335(25) &  0.326(2)	&  0.333(1) &\\
           &   1.0 & 	0.239(6)  &  0.215(2)	&  0.239(3) &\\
	   &   2.0 &	0.168(5)  &  0.118(3)	&  0.166(4) &\\
\hline
1.0	   &   0.0 &	0.873(41) & 0.86767(5)  &  0.8764(3)& 0.8745(2) \\
	   &   0.4 &	0.624(17) &  0.586(4)	&  0.62(2)  & 0.61(1) \\
       	   &   1.0 & 	0.436(11) &  0.384(3)	&  0.43(1)  & 0.44(1) \\
	   &   2.0 &	0.335(45) &  0.205(2)	&  0.30(1)  & 0.315(6)\\
\hline
2.0	   &   0.0 & 	1.547(2)  & 1.54243(9)  &  1.5487(3)& 1.5501(5) \\
	   &   0.4 &	1.10(8)   &  1.045(20)	&  1.08(1)  & 1.10(1) \\
           &   1.0 & 	0.795(25) &  0.721(5)	&  0.780(9) & 0.788(6)  \\
	   &   2.0 &	0.65(8)	  &  0.43(5)	&  0.532(9) & 0.54(1) \\
\hline\hline
\end{tabular}
\end{center}
\end{table}

Let us now compare the tetramer and the full-monomer results for generic values
of $\phi_p$ and vanishing colloidal density. For this purpose we consider the 
solvation properties of a single colloid in the polymer solution. The relevant
quantity here is the insertion free energy, which gives the free energy change 
due to the insertion of a colloid at fixed polymer chemical potential.
Equivalently, one can use the depletion thickness
$\delta_s$, which represents the 
average width of the depleted layer around the colloid.
\cite{FT-07,FST-07,FT-08,LT-11}

\begin{figure}[t]
\begin{center}
\begin{tabular}{c}
\epsfig{file=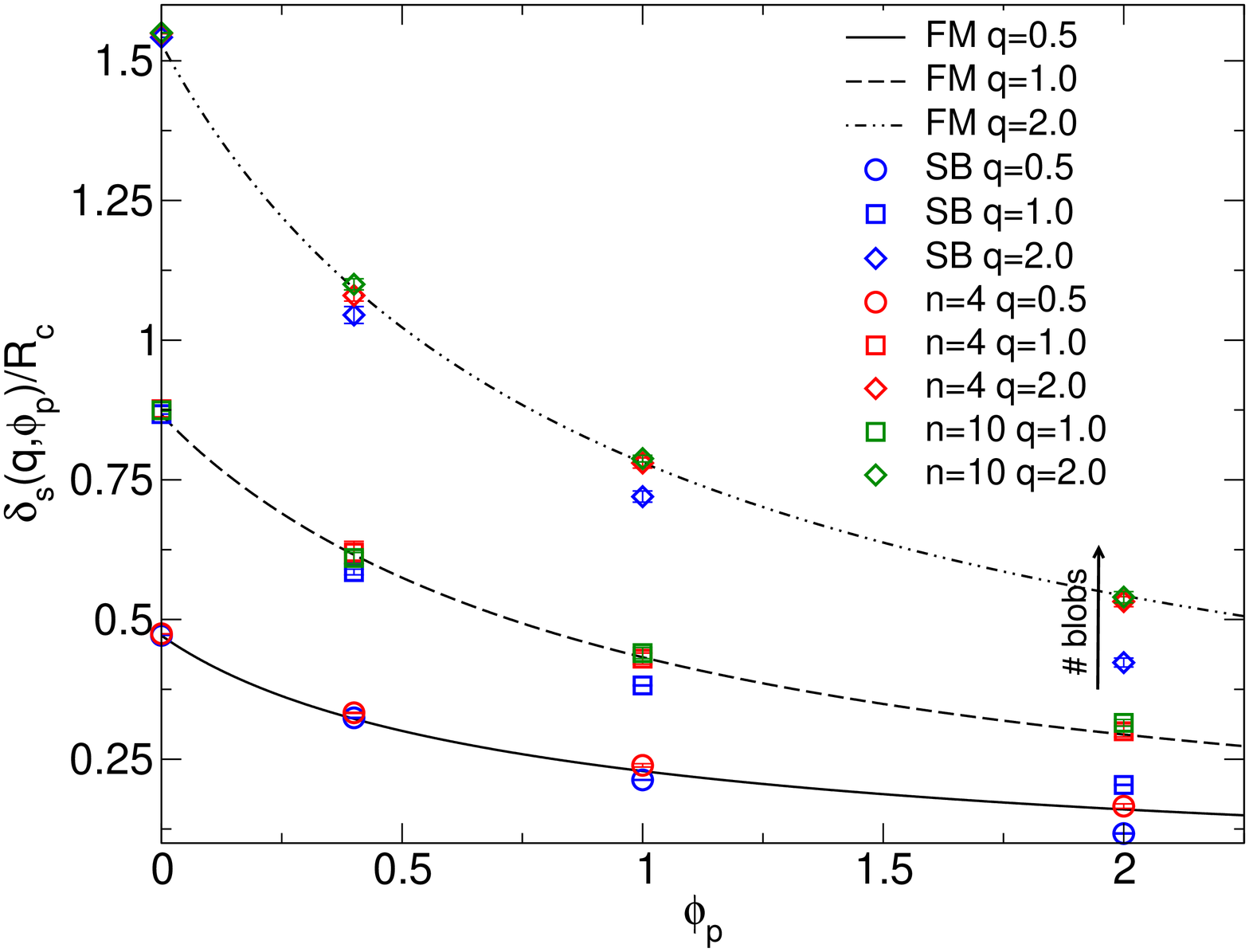,angle=0,width=8truecm} \hspace{0.5truecm} \\
\end{tabular}
\end{center}
\caption{Depletion thickness versus $\phi_p$. 
Full-monomer (lines, FM, from Ref.~\protect\onlinecite{DPP-13-depletion}), 
single-blob (SB), 
tetramer ($n=4$), and decamer ($n=10$) results.
}
\label{fig:depletion-vs-phip}
\end{figure}

Such a quantity can be related to the integral of any polymer-colloid
distribution function. For instance, if $g_{{\rm mon},cp}({\bf r};\mu_p)$ and 
$g_{cp,i}({\bf r};\mu_p)$ are the monomer-colloid and blob-colloid distribution
functions at a given polymer chemical potential $\mu_p$, the integral
$G_{cp}(\mu_p)$ defined by
\begin{eqnarray}
G_{cp}(\mu_p) &=& \int d{\bf r}\, [g_{{\rm mon},cp}(r;\mu_p) - 1]
\nonumber \\
&=& \int d{\bf r}\, [g_{cp,i}(r;\mu_p) - 1],
\label{Gmon}
\end{eqnarray}
is the same for both correlation functions and directly related to the 
insertion free energy.\cite{HMD-06,DPP-13-depletion} 
The depletion thickness is then defined as 
\begin{eqnarray}
\frac{4\pi}{3}\left(R_c+\delta_s\right)^3 = - G_{cp},
\end{eqnarray}
from which it follows
\begin{eqnarray}
{\delta_s\over R_c} = \left(- {{G}_{cp}\over V_c}\right)^{1/3} - 1,
\label{dsoverRcvsGcp}
\end{eqnarray}
where $V_c = 4 \pi R_c^3/3$ is the volume of the colloid.
The depletion thickness was determined for polymer systems and for single-blob
models in Ref.~\onlinecite{DPP-13-depletion}. Here, we extend the calculation
to the CG blob model. Results are reported in 
Table~\ref{tab:deplection-vs-phip} and summarized in 
Fig.~\ref{fig:depletion-vs-phip}. For $q=0.5$ and $q=1$, tetramer and 
full-monomer results are in full agreement up to $\phi_p= 2$.
For $q=2$ the tetramer slightly underestimates the depletion thickness.
Nonetheless, it represents a significant improvement with respect to the 
single-blob model, which becomes increasingly inaccurate as $\phi_p$ 
increases. Again, this is not surprising as we expect the single-blob
model to be reliable only for $q\lesssim 1$.

\section{Thermodynamic properties in the homogeneous phase}
\label{sec5}

\begin{table*}
\caption{Partial structure factors $S_{pp,0}$, $S_{cp,0}$, and $S_{cc,0}$
in the zero-momentum limit (see App.~\ref{Appendix-KB} for the 
definitions) for a few values of $\phi_c = 4\pi R^3_c \rho_c/3$ and 
$\phi_p = 4\pi \hat{R}^3_g \rho_p/3$. 
We report full-monomer (FM) and GFVT results and estimates
for the CG models with $n=1$, 4, and 10 blobs. 
All estimates, except the GFVT results, are appropriate for $L=600$ Domb-Joyce
chains, as explained in the text.
}
\label{table-Salphabeta}
\begin{tabular}{ccccccccc}
\hline\hline
& $q$ & $\phi_c$ & $\phi_p$ & FM & GFVT & $n=1$ & $n=4$ & $n=10$ \\
\hline
$S_{pp,0}$ & 1 & 0.1 & 0.6 &
   1.46(4) & 2.35 & 1.25(2)& 1.39(4) & 1.41(4)\\
            &   & 0.2 & 0.4 &
   2.68(7) & 12.2 & 2.00(3)& 2.62(5) & 2.63(3)\\
            &   & 0.3 & 0.2 &
   2.37(9) &1.72 & 1.69(3)& 2.25(5) & 2.47(5)\\
            & 2 & 0.1 & 1.0 &
   0.71(4) & 1.47 & 0.522(2)& 0.706(20) & 0.71(2)\\
            &   & 0.2 & 0.8 &
   2.6(3) & 1.43 & 0.701(8)& 1.55(9) & 2.6(1)\\
            &   & 0.3 & 0.2 &
   1.5(1) & 0.48 & 0.863(4)& 1.38(4) & 1.58(3)\\
\hline
$S_{cp,0}$ & 1 & 0.1 & 0.6 &
  $-$1.20(4)& $-$2.03 &$-$1.00(2)& $-$1.13(4) & $-$1.16(4)\\
            &   & 0.2 & 0.4 &
  $-$1.49(4)& $-$7.56 &$-$1.07(2)& $-$1.46(30)& $-$1.48(2)\\
            &   & 0.3 & 0.2 &
  $-$0.83(3)& $-$0.65 &$-$0.534(1)& $-$0.785(20) & $-$0.87(2)\\
            & 2 & 0.1 & 1.0 &
  $-$0.84(5)& $-$2.00 &$-$0.516(4)& $-$0.796(25) & $-$0.83(2)\\
            &   & 0.2 & 0.8 &
  $-$2.2(3) & $-$1.33 &$-$0.472(8)& $-$1.25(8)   & $-$2.2(1)\\
            &   & 0.3 & 0.2 &
  $-$0.60(4)& $-$0.147&$-$0.197(1)& $-$0.50(2)   & $-$0.62(1)\\
\hline
$S_{cc,0} $ & 1 & 0.1 & 0.6 &
   1.22(4)& 1.985 & 1.06(2)& 1.15(4) & 1.19(4)\\ 
            &   & 0.2 & 0.4 &
   0.95(2)& 4.80 & 0.70(1)& 0.92(2) & 0.94(1)\\
            &   & 0.3 & 0.2 &
   0.36(1)& 0.32 & 0.247(4)& 0.341(8)& 0.375(8)\\
            & 2 & 0.1 & 1.0 &
   1.28(6)& 3.07 & 0.895(6)& 1.207(35) & 1.27(3)\\
            &   & 0.2 & 0.8 &
   2.0(2)& 1.39 & 0.510(9)& 1.15(7) & 2.0(1)\\
            &   & 0.3 & 0.2 &
   0.319(15)& 0.136 & 0.143(2)& 0.263(7) & 0.326(5)\\
\hline\hline
\end{tabular}
\end{table*}

To quantify the accuracy of the CG procedure,
we wish now to compare the predictions of the tetramer model 
with results obtained in full-monomer simulations for finite 
volume fractions $\phi_c$
and $\phi_p$. This is not an easy task. Indeed, since the target 
functions used to determine the CG potentials were computed in the 
scaling limit $L\to\infty$, the tetramer model provides results 
that can be considered asymptotic.
Therefore, a meaningful 
comparison requires also an extrapolation of the full-monomer results 
to the scaling limit, which is too demanding from the computational
point of view. To avoid any extrapolation we have decided to take 
a slightly different approach. Instead of considering 
a CG tetramer model that reproduces the scaling behavior of the
polymer system, we consider a CG model that is appropriate 
to describe Domb-Joyce chains with $L=600$ monomers, which are taken
as reference system. Then, it makes sense to compare CG results with 
full-monomer simulations of $L=600$ chains at finite density, without
performing any extrapolation. Of course, the price to be paid is that 
we need to recompute all CG potentials taking the distribution functions
computed with $L=600$ chains as targets.
In the pure polymer case, results for $L=600$ are 
essentially already in the scaling limit, hence there is no need 
to recompute the intramolecular tetramer potentials
and the blob-blob intermolecular potentials. Differences 
of the order of a few percent are instead
observed for the blob-colloid distribution functions. We have therefore
recomputed the corresponding potentials.
A comparison of $L=600$ and scaling data is presented in
the supplementary material:\cite{suppl} differences are small,
but not negligible.

In order to compare the thermodynamic behavior, 
we focus on the zero-momentum limit of the partial structure factors
$S_{\alpha\beta,0}$ 
(Kirkwood-Buff integrals), which 
can be related to several thermodynamic quantities.\cite{KB-51,BenNaim}
They are computed as discussed in App.~\ref{Appendix-KB}. 
In Table~\ref{table-Salphabeta} we report their
estimates for several values of $\phi_c$ and $\phi_p$, which are close to the 
binodal, as predicted by GFVT.

\begin{figure}[tb!]
\begin{center}
\begin{tabular}{c}
\epsfig{file=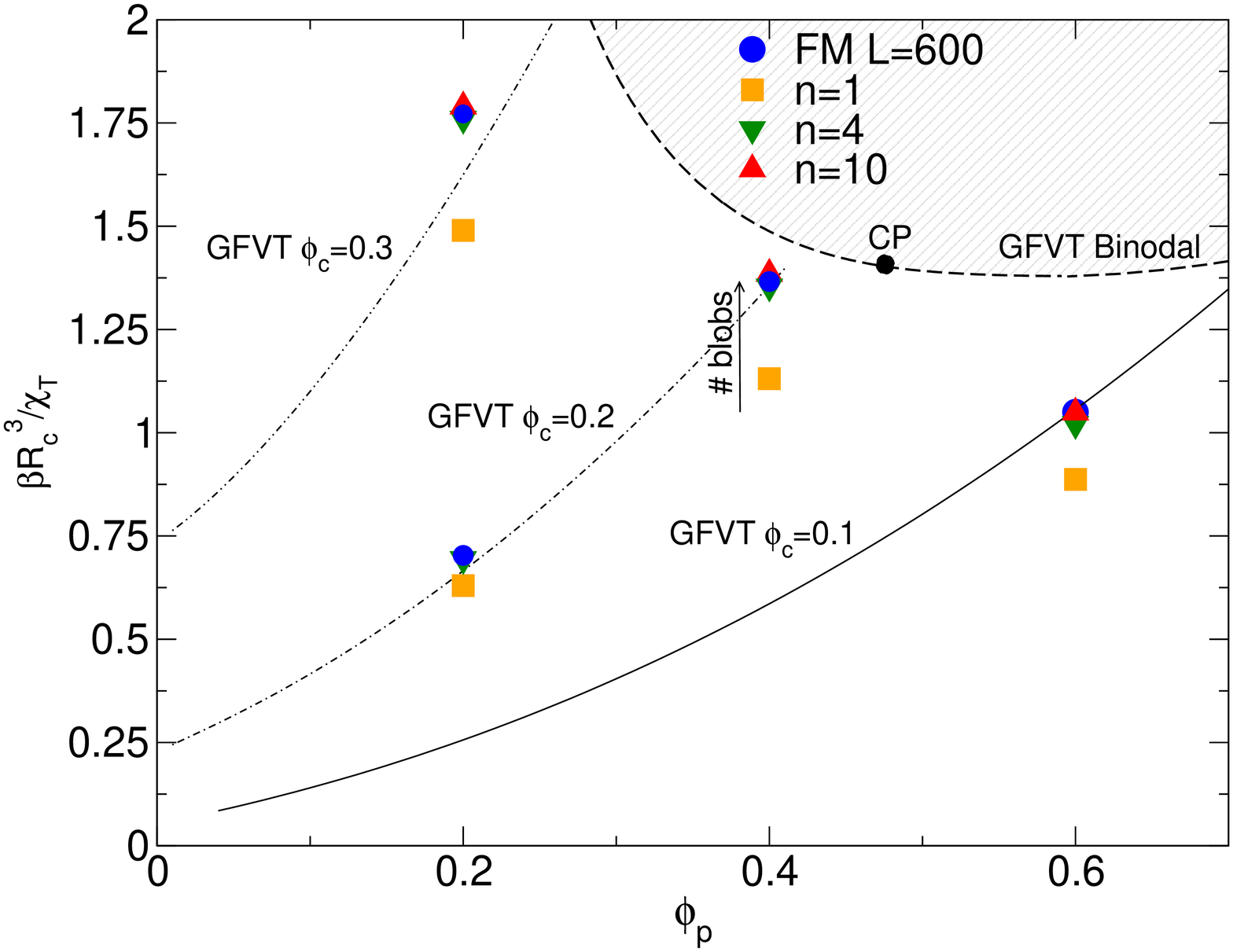,angle=0,width=8truecm} \hspace{0.5truecm}\\
\epsfig{file=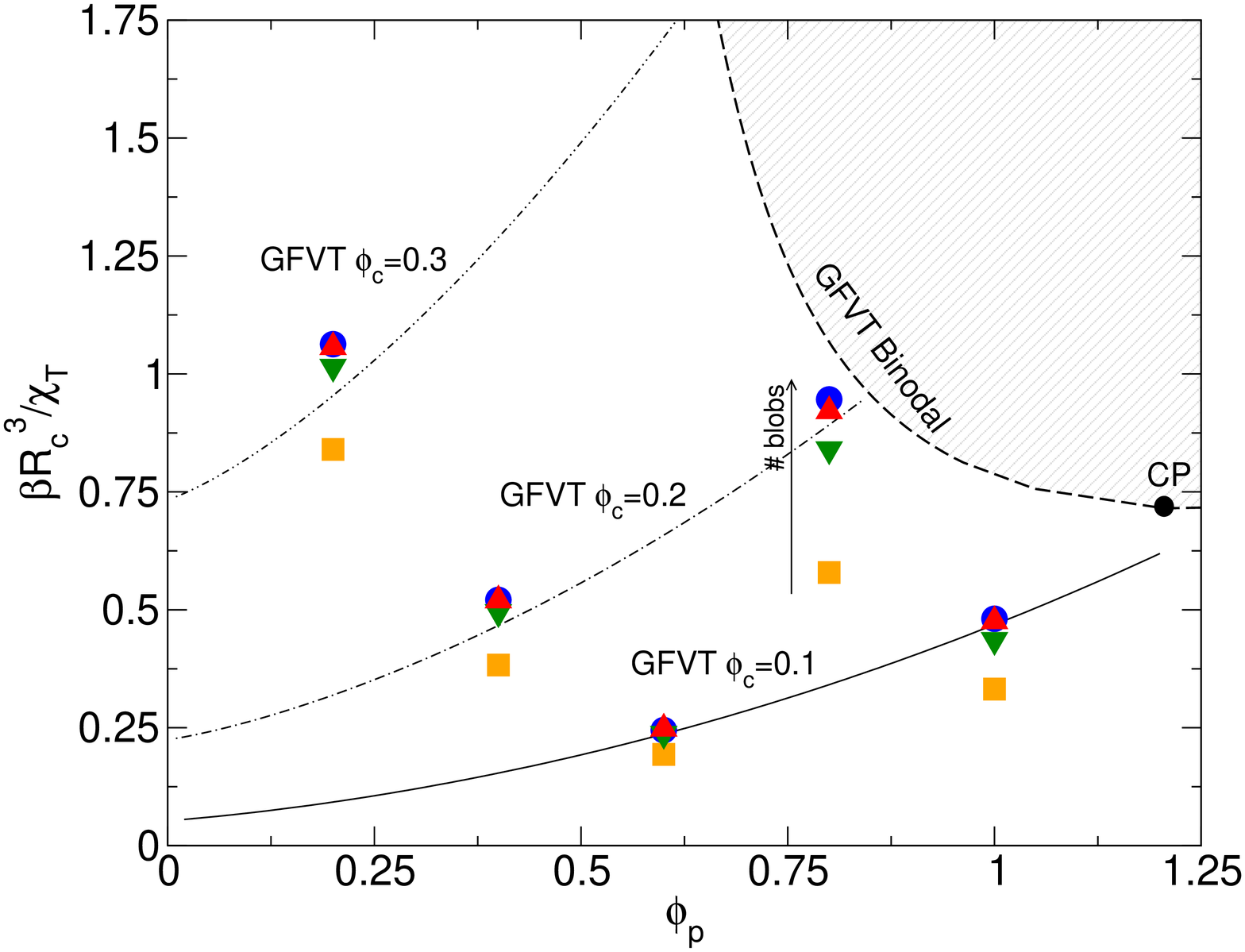,angle=0,width=8truecm} \hspace{0.5truecm}\\
\end{tabular}
\end{center}
\caption{Inverse isothermal compressibility as a function of $\phi_p$ for 
three values of $\phi_c$, 0.1, 0.2, and 0.3. Lines give the GFVT 
prediction, points are simulation results obtained by using DJ 
L=600 chains (FM) and the corresponding CG models with $n=1$,
4  and 10 blobs. We also report (dashed line in the 
upper right part of each figure) the GFVT binodal (note that 
in the GFVT approximation $\chi_T$ is not singular at the critical
point) and the corresponding critical points (CP). Top panel refers to
$q=1$, bottom panel to $q=2$.
}
\label{fig:kappaT}
\end{figure}

In general, we find that the estimates of the partial
structure factors $|S_{\alpha\beta,0}|$ (see App.~\ref{Appendix-KB} for 
definitions)
increase with the number $n$ of blobs towards the corresponding full-monomer value: 
$|S_{\alpha\beta,0}(n=1)| < |S_{\alpha\beta,0}(n=4)|
\lesssim |S_{\alpha\beta,0}(\hbox{FM})|$. 
The single-blob model always underestimates $|S_{\alpha\beta,0}|$:
the relative difference with the full-monomer estimates increases
somewhat with $\phi_c$ and especially with $q$. For $q=2$ the model 
is clearly unreliable. 
For $q = 1$, the tetramer model gives results that are consistent
with the full-monomer ones within errors, representing a 
significant improvement with respect to the single-blob model.
For $q=2$, tetramer results are also close to the full-monomer ones
for $\phi_c = 0.1$. However, for $\phi_c = 0.2$ and $0.3$, the 
zero-momentum structure factors are sligthly underestimated, confirming 
that a higher-resolution model is needed to correctly
model polymer-colloid mixtures for $q=2$.

In Table~\ref{table-Salphabeta} we also report GFVT results.
Since they are obtained by using scaling-limit expressions for the 
depletion thickness,\cite{DPP-13-depletion,DPP-14-GFVT} 
a small correction should be applied to the GFVT results, before 
comparing them with the CG and FM ones, appropriate for $L=600$ Domb-Joyce
chains. Such correction, however, is of the order of a few percent, hence
small compared with the differences we observe.
For $q=1$ and $\phi_c = 0.1$, 0.2, GFVT significantly overestimates 
the structure factors. This is particularly evident for $\phi_c = 0.2$,
$\phi_p = 0.4$, which is very close to the GFVT critical point 
$\phi_{c,\rm crit} \approx 0.18$,
$\phi_{p,\rm crit} \approx 0.47$, but which is far from 
the full-monomer critical point \cite{DPP-14-GFVT,MLP-12}
$\phi_{c,\rm crit} \approx 0.22$,
$\phi_{p,\rm crit} \approx 0.62$.
For $q=2$, $|S_{\alpha\beta,0}|$ is overestimated for 
$\phi_c = 0.1$, while it is underestimated by a factor of 2-3
for $\phi_c = 0.2$ and 0.3. The large value for $\phi_c = 0.1$ 
is a direct consequence of the nearby presence of  the critical 
point (GFVT predicts 
$\phi_{c,\rm crit} \approx 0.11$,
$\phi_{p,\rm crit} \approx 1.21$), not confirmed by the 
full-monomer data, that are instead consistent with the 
estimate\cite{DPP-14-GFVT,MLP-12}
$\phi_{c,\rm crit} \approx 0.19$,
$\phi_{p,\rm crit} \approx 1.08$.

\begin{figure}[tb!]
\begin{center}
\begin{tabular}{c}
\includegraphics[width=8.5cm]{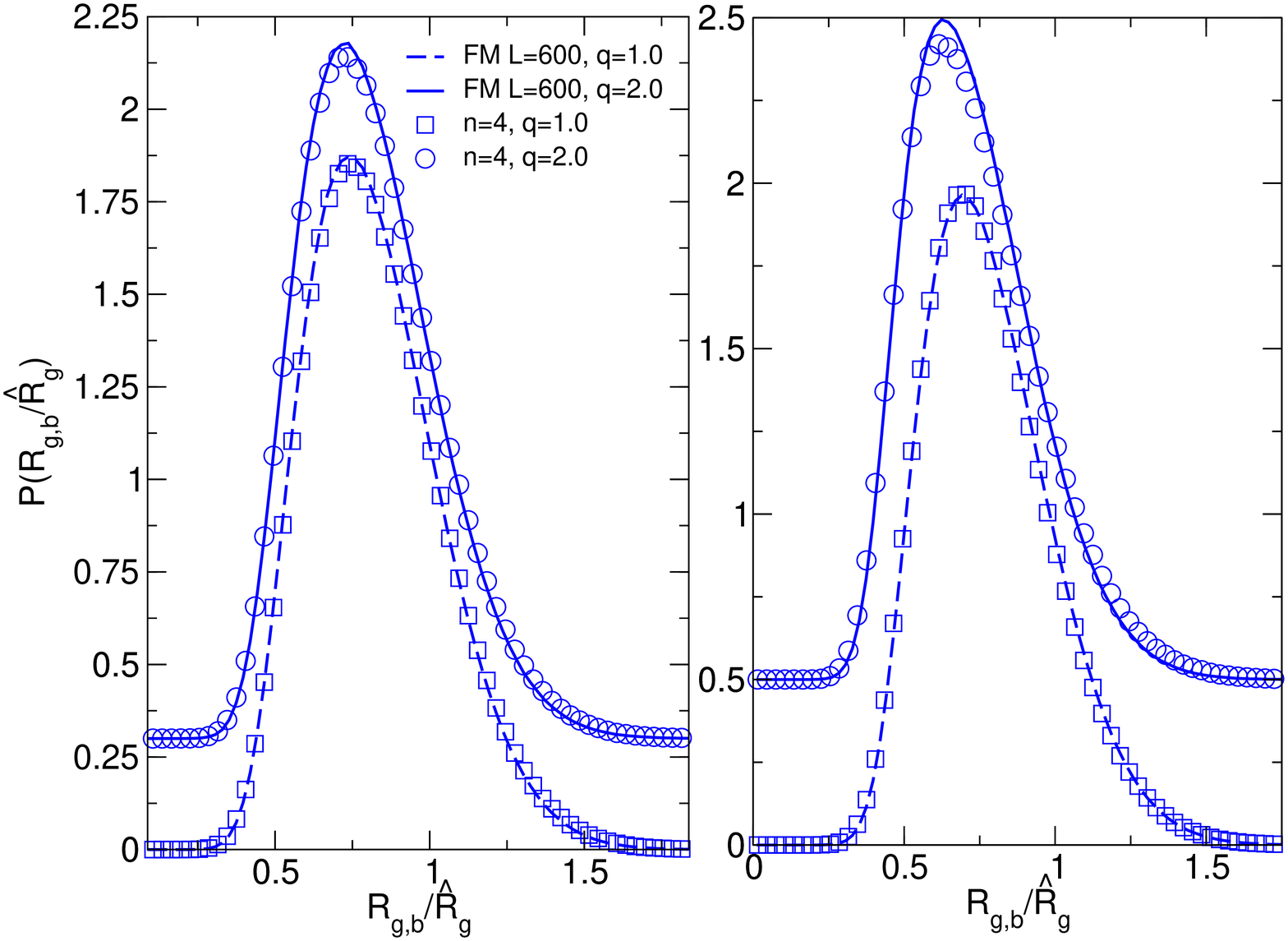} \\
\end{tabular}
\end{center}
\caption{Distribution functions of the radius of gyration $R_{gb}$ of the CGR
of the polymers for the tetramer.
We report data for 
$\phi_c=0.1$, $\phi_p=0.6$ (left) and 
$\phi_c=0.3$, $\phi_p=0.2$ (right).
The data for $q=2$ are shifted upward by 0.3 (left) and 0.5 (right) for clarity.
If we average $R_{g,b}^2$ over the distribution, we obtain 
$\langle R_{g,b}^2 \rangle^{1/2}/\hat{R}_g \approx 0.85$ ($q=1$, $\phi_c=0.1$,
$\phi_p=0.6$), 0.81 ($q=1$, $\phi_c=0.3$, $\phi_p=0.2$),
0.83 ($q=2$, $\phi_c=0.1$, $\phi_p=0.6$), and 
0.76 ($q=2$, $\phi_c=0.3$, $\phi_p=0.2$).
}
\label{fig:rg2-finite-density}
\end{figure}

\begin{figure}[tb!]
\begin{center}
\begin{tabular}{c}
\includegraphics[width=7.5cm]{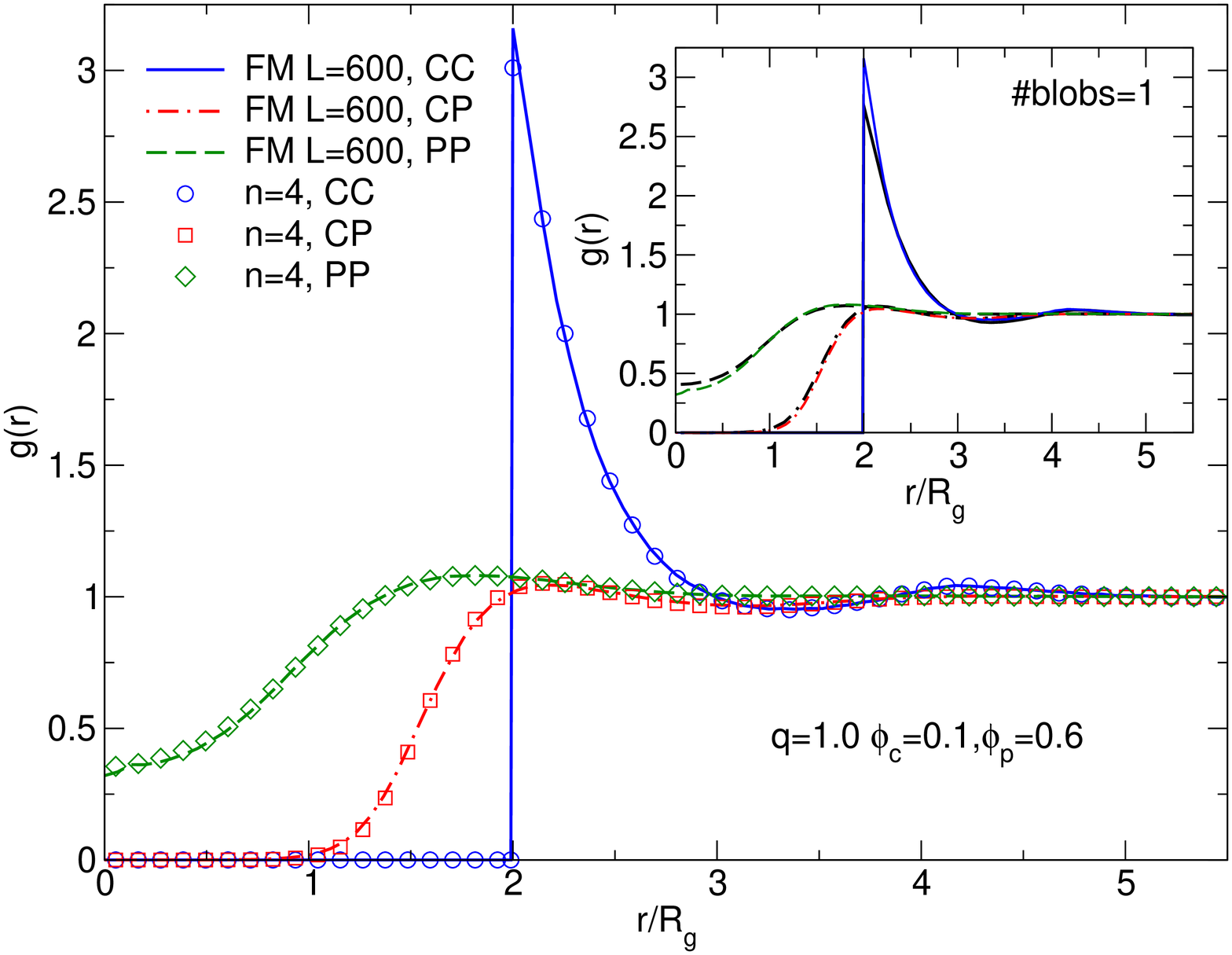} \\
\includegraphics[width=7.5cm]{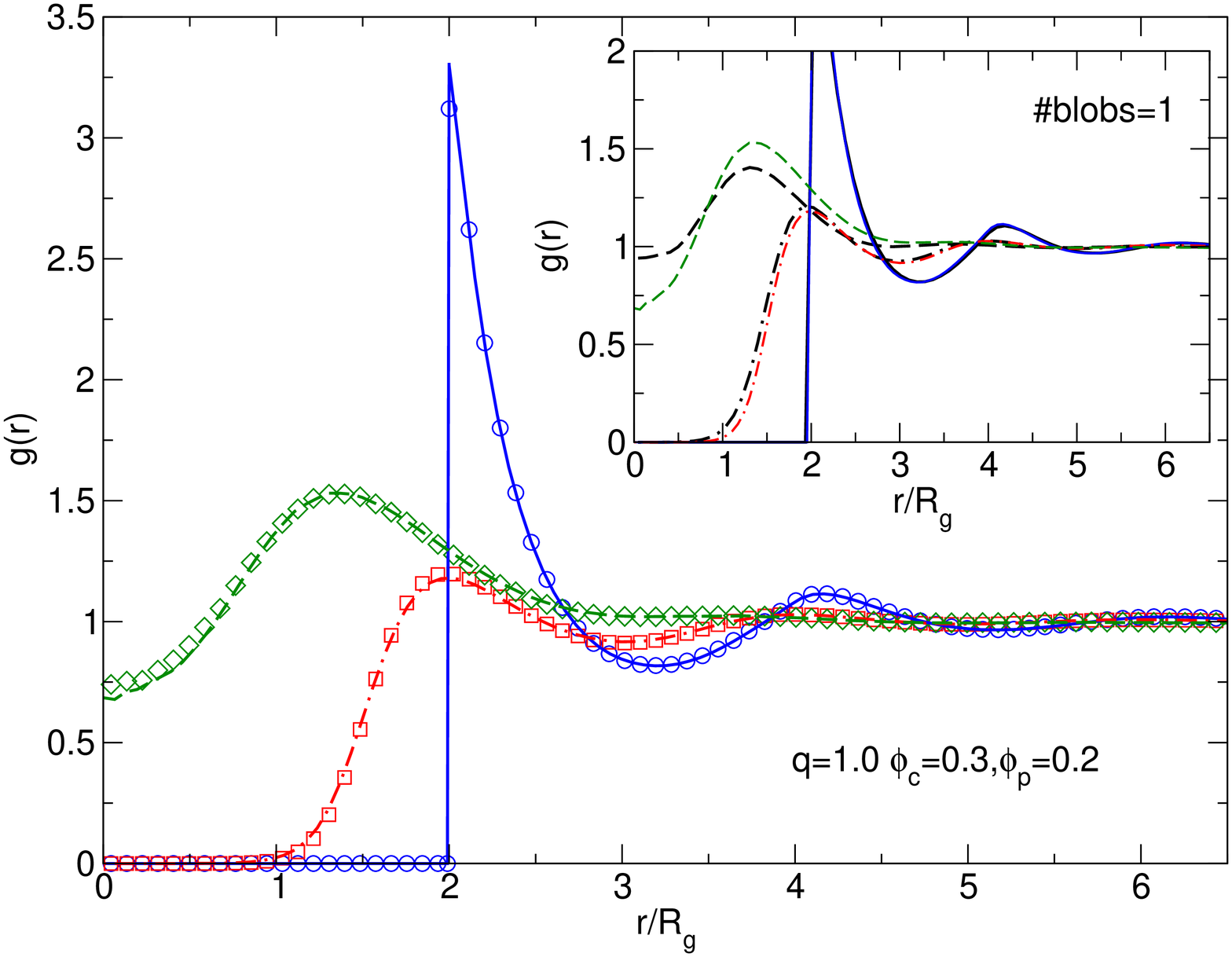} \\
\end{tabular}
\end{center}
\caption{Pair distribution functions between the 
centers of mass of the molecules as a function of $r/\hat{R}_g$: 
$g_{cc}$, $g_{cp}$, and $g_{pp}$ are the colloid-colloid, colloid-polymer,
and polymer-polymer functions, respectively. We report full-monomer (FM)
$L=600$ Domb-Joyce estimates and the corresponding 
$n=1$ (inset), and $n=4$ results for $q=1$. 
We report results for 
$\phi_c = 0.1$, $\phi_p = 0.6$ (top), and for 
$\phi_c = 0.3$, $\phi_p = 0.2$ (bottom).
}
\label{fig:gr-homogeneous-q1}
\end{figure}

\begin{figure}[tb!]
\begin{center}
\begin{tabular}{c}
\includegraphics[width=7.5cm]{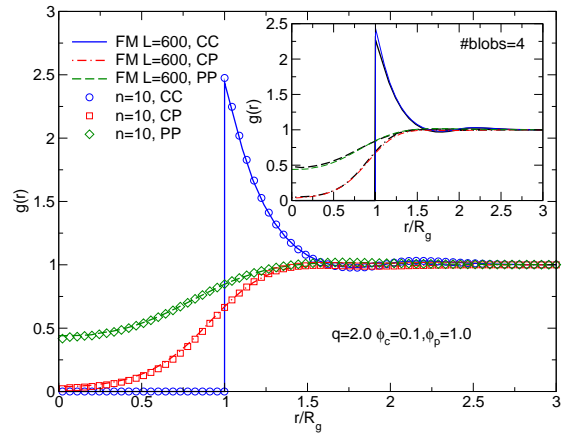} \\
\includegraphics[width=7.5cm]{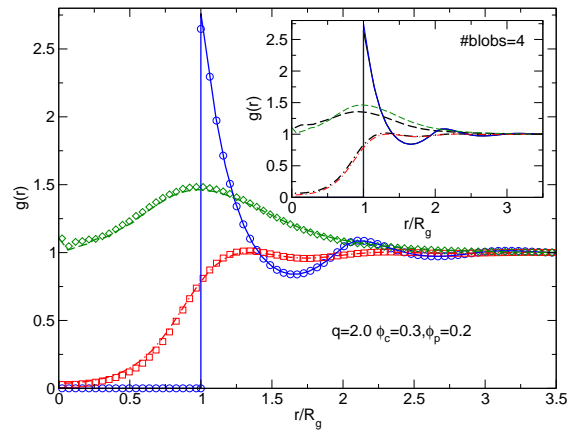} \\
\end{tabular}
\end{center}
\caption{Pair distribution functions between the 
centers of mass of the molecules as a function of $r/\hat{R}_g$: 
$g_{cc}$, $g_{cp}$, and $g_{pp}$ are the colloid-colloid, colloid-polymer,
and polymer-polymer functions, respectively. We report full-monomer (FM)
$L=600$ Domb-Joyce estimates and the corresponding 
$n=4$ (inset) and $n=10$ results for $q=2$. 
We report results for $\phi_c = 0.1$, $\phi_p = 1.0$ (top) and for 
$\phi_c = 0.3$, $\phi_p = 0.2$ (bottom).
}
\label{fig:gr-homogeneous-q2}
\end{figure}

As an additional check, we have computed several thermodynamic quantities
that can be obtained starting from the zero-momentum partial structure
factors, 
see appendix~\ref{Appendix-KB} for definitions 
and the supplementary material for an extensive list of results. 
In Fig.~\ref{fig:kappaT} we report $\beta R_c^3/\chi_T$, where 
$\chi_T$ is the isothermal compressibility. For both $q=1$ and $q=2$ 
single-blob results are clearly unreliable, 
discrepancies increasing with $q$ and
$\phi_p$. Tetramer results represents a significant improvement. 
They fall on top of the full-monomer estimates for $q=1$, 
while for $q=2$ small differences
can still be seen for $\phi_c = 0.2$ and $\phi_c = 0.3$. GFVT appears quite 
reliable for this quantity: unexpectedly, 
the discrepancies observed for $S_{\alpha\beta,0}$
cancel out in this combination. However, significant
discrepancies are expected at the critical point. Indeed, 
at the critical point $\chi_T$ diverges. Instead,
in the GFVT approximation 
$\chi_T$ is finite at criticality (GFVT is essentially a mean-field theory).
Thus, it is not surprising that this approximation is not accurate for the 
critical point position.\cite{DPP-14-GFVT}

Let us now consider the intramolecular polymer structure. We have verified 
that the 
tetramer model reproduces well the blob-blob pair distribution function
computed by using the CGR of the polymer system. As an additional
check, we consider the radius of gyration of the CGR of the polymers, 
defined as 
\begin{equation}
   R^2_{g,b} = {1\over 2 n^2} \sum_{i=1}^n ({\bf s}_i - {\bf s}_j)^2.
\end{equation}
Its distribution for $n=4$ is compared with that of the radius of gyration of 
the tetramers in Fig.~\ref{fig:rg2-finite-density}. The agreement is excellent, 
confirming the 
accuracy of the tetramer representation. Note also that polymers become more compact 
as $q$ and/or $\phi_c$ increases: the distributions indeed move to the left 
as these two quantities become larger (additional comments and data are reported in
App.~\ref{App.B}).

Finally, we consider the intermolecular structure, comparing 
the center-of-mass distribution functions.
Results are reported in Figs.~\ref{fig:gr-homogeneous-q1} and 
\ref{fig:gr-homogeneous-q2}. For $q=1$, the tetramer model reproduces 
quite well the 
full-monomer results. For $q=2$, $g_{cc}(b,q)$ and $g_{cp}(b,q)$
are well reproduced. On the other hand, significant differences are 
observed for the polymer-polymer pair distribution function for 
$\phi_c = 0.3$, $\phi_p = 0.2$. This behavior is quite general: For all
CG systems discrepancies always increase as $\phi_c$ is increased. This is not surprising,
since the role of the neglected many-body colloid-polymer interactions increases as 
the colloid density gets larger.

\section{Higher-resolution models and transferability} \label{sec6}

\begin{table*}[tbp!]
\caption{Virial combinations for full-monomer Domb-Joyce walks with 
$L=600$ (FM) and for three different versions of the decamer model,
as explained in the text. Symbols are the same as in Table 
\protect\ref{tab:virial-tetramer}.}
\label{tab:virial-decamer}
\begin{center}
\begin{tabular}{ccccccccc}
\hline
\hline
$q$ & model & $A_{2,cp}$ & $A^I_{3,cpp}$ & $A^{\rm fl}_{3,cpp}$ & $A_{3,cpp}$  
                 & $A^I_{3,ccp}$ & $A^{\rm fl}_{3,ccp}$ & $A_{3,ccp}$ \\
\hline
$1$ & 
FM & 26.77(1) & 134.7(2) & $-$5.8(1) & 128.9(2) & 
                355.2(4) & $-$10.8(1) & 345.2(4) \\
& (a) & 26.014(7)& 134.3(3) & $-$6.0(2) & 127.04(7) & 
                347.1(6) & $-$11.4(4) & 335.7(6)\\
& (b) & 27.386(7) & 141.7(3) & $-$6.0(2) & 135.8(4) &  
                369.2(6) & $-$11.4(4) & 357.8(6)\\
& (c) & 26.920(7) & 138.5(3) & $-$6.0(2) & 132.4(4) & 
                360.6(6) & $-$11.0(4) & 349.6(6)\\
\hline
2 & 
FM &8.234(5) & 26.13(3) & $-$1.67(2) & 24.45(5) &  
               16.60(2) & $-$0.87(1) & 15.73(2) \\
& (a) & 7.882(3) & 25.2(1) & $-$1.6(1) & 23.6(1) & 
               16.14(8) & $-$0.88(4) & 15.26(9) \\
& (b) & 8.355(3) & 26.9(1) & $-$1.7(1) & 25.2(1) & 
               17.00(8) & $-$0.92(4) & 16.08(9) \\
& (c) & 8.298(3) & 26.7(1) & $-$1.6(1) & 25.0(1) & 
               16.85(9) & $-$0.91(4) & 15.94(9) \\
\hline\hline
\end{tabular}
\end{center}
\end{table*}

The results presented in 
 Sec.~\ref{sec5}, show that the CG model becomes inaccurate 
along the binodal when the blob size is comparable with the 
colloid radius. Indeed, for the single-blob model, differences are observed 
for $q=1$, i.e.~when $\hat{R}_g$, which is the size of the blob,  is 
equal to $R_c$. 
The tetramer begins to break down for $q=2$, which corresponds to 
$\hat{r}_g/R_c \approx 0.9$
(for $n=4$ we have \cite{DPP-12-Soft} $\hat{r}_g/\hat{R}_g \approx 0.44$, where
$\hat{r}_g$ is the blob zero-density radius of gyration). 
Again, blob size and colloid
radius are comparable.
Therefore, in order to study
polymer-colloid systems in the protein regime $q>1$, we need to develop
higher-resolution models with a larger number of blobs per chain.
Since \cite{DPP-12-Soft} $\hat{r}_g/\hat{R}_g \approx n^{-\nu}$, 
studies of mixtures with size ratio $q$ require CG systems with at 
least $n = q^{1/\nu}$ blobs. 

To derive $n$ blob models,
one could address the problem directly, measuring 
the intermolecular blob-colloid correlation functions for a CGR
of the polymer with $n$ blobs and determine the corresponding $n/2$ 
potentials by using the IBI method. This is probably feasible for $n$ not 
too large. Here, however, we will use a simpler approach
based on the idea of the transferability of the interactions, which has been
shown to work nicely for pure polymer systems, both under good-solvent 
conditions \cite{DPP-12-JCP} and in the thermal crossover region.
\cite{DPP-13-thermal} Such an approach is based on the assumption
that the potentials are independent of the model resolution once the blob
radius of gyration $\hat{r}_g$ is used as reference length scale. 
In the presence of the colloids, we should also take into account 
a second length scale, the colloid radius $R_c$. If only
pair interactions are relevant, it is natural to assume that the 
blob-colloid interactions depend only on the ratio $q_b = \hat{r}_g/R_c$ 
between the radius of gyration of the blob and $R_c$: the colloid-blob
interaction is the same for systems with different resolutions but with the 
same $q_b$. If this assumption holds, we can transfer the tetramer potentials
to higher-resolution CG systems. The two assumptions are essentially based on the 
idea that polymers are self-similar objects, so that each subchain has the 
same structure as the full polymer in the scaling limit.

In practice, let us indicate with $V_{cp,i}(b,q)$ the
blob-colloid potential for the tetramer; here, $i$ labels the blob along
the chain and $b = r/\hat{R}_g$, where $\hat{R}_g$ is the 
zero-density polymer radius of gyration. Assuming transferability,
we set for the potentials for the $n$-blob model
[model (a)]:
\begin{eqnarray}
&& V_{cp,1} (b,q; n) = V_{cp,n}(b,q;n) = V_{cp,1} (\lambda_n b,
q/\lambda_n)
\\
&& V_{cp,i} (b,q; n) = V_{cp,2} (\lambda_n b,q/\lambda_n)
\qquad 2 \le i \le n-1. \nonumber 
\end{eqnarray}
Here $\lambda_n$ is given by
\begin{equation}
\lambda_n = {\hat{r}_g(4)\over \hat{r}_g(n)},
\label{def-lambdan}
\end{equation}
where $\hat{r}_g(n)$ is the average radius of gyration of the blob in 
the $n$-blob CGR of the polymer (numerical results are 
reported in App.~A of Ref.~\onlinecite{DPP-12-Soft}).

To verify the quality of the approximation we have considered the decamer 
model with $n=10$ blobs. To derive the potentials for our reference values 
$q=1$ and $q=2$, we need to derive first the tetramer potentials for the 
corresponding ratios $q/\lambda_{10}$. Since \cite{DPP-12-Soft}
$\lambda_{10} = 1.702$, we have repeated the determination of the tetramer 
model for $q = 0.587$ and $q = 1.175$.
To avoid uncertainties due to the scaling approximation we have first
used the tetramer model appropriate
for $L=600$ Domb-Joyce walks. 

As a first test of the decamer model, 
we determined the virial combinations
$A_{2,cp}$, $A_{3,ccp}$, and $A_{3,cpp}$. 
Results, labelled (a), are reported in
Table~\ref{tab:virial-decamer}. The model works quite well.
For $q=1$ the predicted $A_{2,cp}$, $A_{3,cpp}$, and $A_{3,ccp}$ 
differ by 3\%, 1\%, and 3\% from the full-monomer data. 
For $q=2$ differences are only slightly larger (4\%, 3\%, 3\%, respectively).

One can surmise that the small differences are end effects, which are expected to 
become progressively irrelevant as the number of blobs increases, 
related to our 
choice of using  $V_{cp,1}(b,q)$ for the two end-blobs and 
$V_{cp,2}(b,q)$ for all internal blobs. To understand the sensitivity of the 
results on this choice, we define a second $n$ blob model by setting
[model (b)]
\begin{equation}
V_{cp,i} (b,q; n) = 
  {1\over 2} [V_{cp,1} (\lambda_n b,q/\lambda_n) + 
              V_{cp,2} (\lambda_n b,q/\lambda_n)],
\label{Vcp-modelb}
\end{equation}
for all $i$. For $n=10$, no significant differences are observed, see 
Table~\ref{tab:virial-decamer}, results labelled (b). Model (b) 
is slightly less accurate than model (a) for $q=1$ and slightly more 
accurate for $q=2$. These comparisons show that the transferability 
hypothesis works quite well, providing us with a model that can be used 
for larger values of $q$ and $\phi_p$ with respect to the tetramer one.

To increase the accuracy of the CG model and obtain estimates of the virial
coefficients that are as precise as the tetramer ones, we now define
a third version [model (c)], in which the potentials are defined as 
in Eq.~(\ref{Vcp-modelb}), but the length rescaling is optimized 
to obtained a better agreement between the full-monomer and the CG estimate
of $A_{2,cp}$. We set therefore [model (c)]
\begin{equation}
V_{cp,i} (b,q; n) = 
  {1\over 2} [V_{cp,1} ({\lambda'}_n b,q/\lambda_n) + 
              V_{cp,2} ({\lambda'}_n b,q/\lambda_n)].
\label{Vcp-modelc}
\end{equation}
In this expression $\lambda_n$ is still given in Eq.~(\ref{def-lambdan}), 
while 
\begin{equation}
{\lambda'}_n = \left({A_{2,cp}(b) \over A_{2,cp}(FM)}\right)^{1/3}
         {\hat{r}_g(4)\over \hat{r}_g(n)},
\end{equation}
where $A_{2,cp}(b)$ and $A_{2,cp}(FM)$ are the estimates obtained 
by using model (b) and the full-monomer model. If polymers were 
monoatomic molecules, model (c) would provide the correct estimate of 
$A_{2,cp}$. In our case, an exact equality does not hold. Still, 
model (c) reproduces the full-monomer value of $A_{2,cp}$ 
with an error of less than 1\%,
which is enough for our purposes. Therefore, in the following we will consider
model (c) for the decamer. 

The analysis reported above was performed using the potentials appropriate 
for Domb-Joyce chains with $L=600$ monomers. We have repeated the calculation 
determining the decamer potentials appropriate to describe polymers in the 
scaling limit. The corresponding virial combinations are reported in 
Table~\ref{tab:virial-tetramer}. We have also recomputed the depletion
thickness, see Table~\ref{tab:deplection-vs-phip} 
and Fig.~\ref{fig:depletion-vs-phip}.
The decamer 
and the tetramer give consistent results for both $q=1$ and $q=2$, indicating 
that both CG models describe accurately the solvation properties of a single 
colloid up to $\phi_p = 2$. 

In the homogeneous phase close to the fluid-fluid binodal, the tetramer model 
is only accurate for $q=1$. For $q=2$ differences are clearly observed for 
$\phi_c = 0.2$ (close to the critical point) and for $\phi_c = 0.3$, 
close to the colloid-liquid phase, see Table~\ref{table-Salphabeta}. 
For these values of $\phi_c$, the tetramer model underestimates
$|S_{\alpha\beta,0}|$. On the other hand, the decamer estimates are 
consistent with the full-monomer ones. Thus, while we expect the 
tetramer to provide the correct phase behavior for polymer-colloid 
mixtures up to $q=1$, for $q=2$ the decamer should be the model of choice.

\begin{figure}[t]
\begin{center}
\begin{tabular}{c}
\epsfig{file=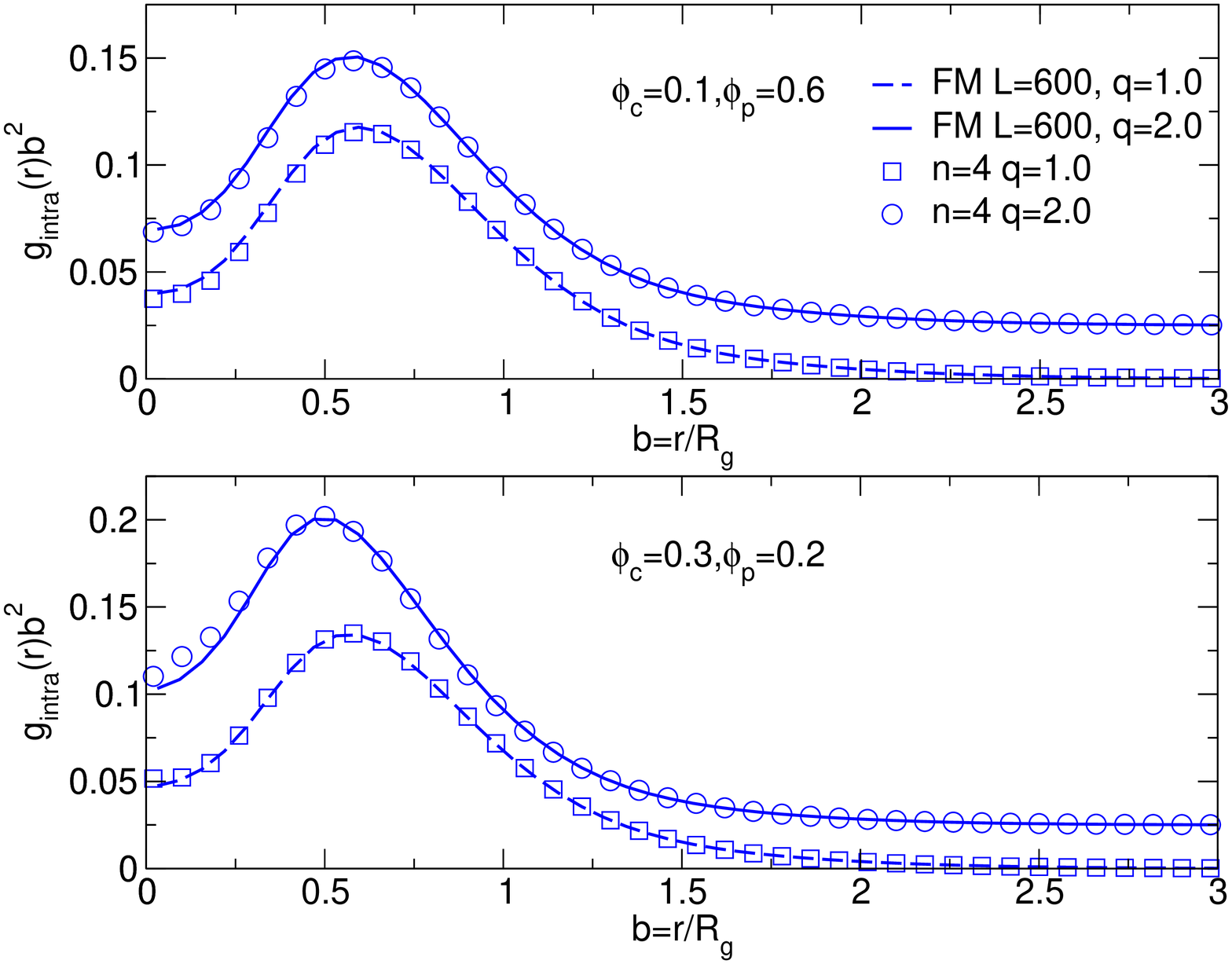,angle=0,width=8truecm} \hspace{0.5truecm} \\
\epsfig{file=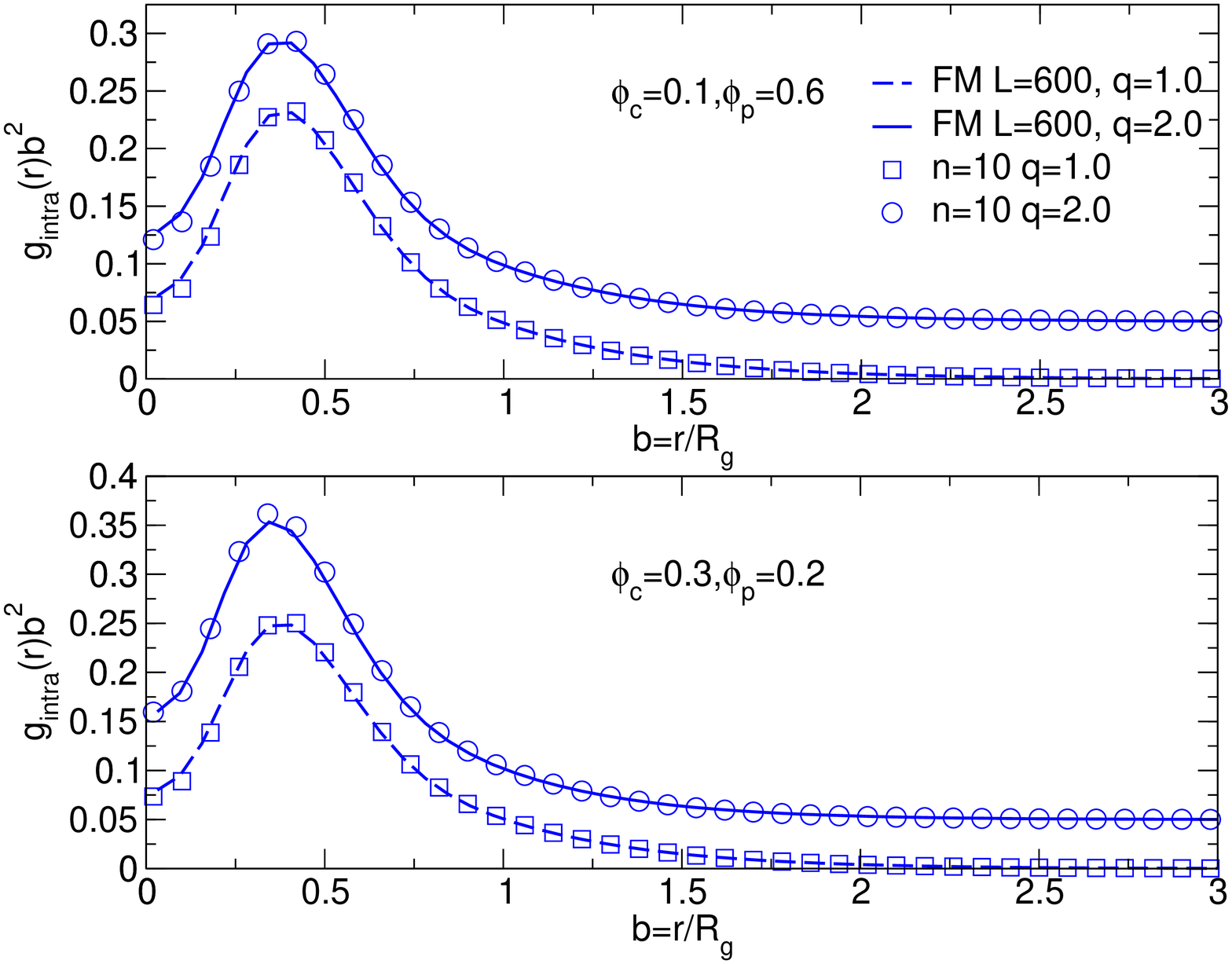,angle=0,width=8truecm} \hspace{0.5truecm} \\
\end{tabular}
\end{center}
\caption{Intramolecular blob-blob distribution function for $n=4$ 
(two upper panels),
and $n=10$ (two lower panels) at two different state points. 
We report results for 
$q=1$ (dashed line and squares) and $q=2$ (continuous line and circles)
versus $b = r/\hat{R}_g$. 
We report full-monomer CGR data (FM, lines) and results for the CG models
(points). Data for $q=2$ have been shifted upward by 0.025 for clarity.
}
\label{fig:gintra-homogeneous}
\end{figure}

We have also verified that the decamer model reproduces the intramolecular and 
intermolecular structure. In Fig.~\ref{fig:gintra-homogeneous} we show
the intramolecular pair distribution function $g_{\rm intra}(b)$ for the 
decamer. It is completely consistent with the blob-blob  
pair distribution function for the CGR of the polymers. 
A similar excellent agreement is observed for the 
$R_{g,b}$ distribution (not shown), 
confirming the good accuracy of the transferability assumption.
Good agreement is also observed for the 
intermolecular structure, see Figs.~\ref{fig:gr-homogeneous-q1} and 
\ref{fig:gr-homogeneous-q2}. In particular, the decamer reproduces 
the polymer-polymer distribution function $g_{pp}(b,q)$ for $q=2$,
$\phi_c = 0.3$, $\phi_p = 0.2$, at variance with 
the tetramer case.

\section{The density-dependent single-blob model} \label{sec7}

As we have discussed, the single-blob model gives a poor description of 
polymer-colloid mixtures for $q\gtrsim 1$, the discrepancies increasing
as the binodal is approached, see Fig.~\ref{fig:kappaT}. Moreover, 
even for polymers, this CG model is not accurate as soon
as $\phi_p\gtrsim 1$. We wish now to consider a variant of the 
single-blob model, proposed in 
Refs.~\onlinecite{LBHM-00,BLHM-01,BLH-01,BLH-02}, which considers interactions
dependent on the polymer density.

For pure polymer systems the method works as follows. One considers a 
thermodynamic state point characterized by a volume fraction $\phi_p$ or,
equivalently, by the excess chemical potential $\mu_p^{({\rm exc})}$ and 
determines the center-of-mass distribution function $g_{pp,FM}(b)$ in 
the polymer (full-monomer) system, where $b = r/\hat{R}_g$. Because
of the equivalence of the ensembles, canonical calculations at $\phi_p$ 
and grand-canonical computations at $\mu_p^{({\rm exc})}$ give the 
the same result for the distribution function in the infinite-volume limit. 
Then, in the spirit
of the structural approach, one determines the potential for the single-blob
model so that the distribution function computed in the CG model is the same 
as the full-monomer counterpart $g_{pp,FM}(b)$. However, as discussed 
in Ref.~\onlinecite{DPP-13-state-dep}, this second step is not defined 
unambiguously,
as the result of the procedure depends on the ensemble.
For instance, one can require the CG model to reproduce $g_{pp,FM}(b)$ in 
the canonical ensemble at volume fraction $\phi_p$. This procedure provides 
a potential $V_{pp,\rm can}(b;\phi_p)$.
Alternatively, one can require the CG
model to reproduce $g_{pp,FM}(b)$ in 
the grand-canonical ensemble at $\mu_p^{({\rm exc})}$. One obtains a 
pair potential
$V_{pp,GC}(b;\mu_p^{({\rm exc})})$, which, however,
 differs from the canonical one.\cite{DPP-13-state-dep} This 
is an intrinsic property of any structural procedure (force-matching
methods do not have this limitation\cite{DPP-13-state-dep}) that 
maps the original system onto a CG system with state-dependent interactions.
\cite{Louis-02,DPP-13-state-dep}

The above strategy
can be directly extended to the mixture. Consider now
a thermodynamic state point with volume fractions $\phi_c$, $\phi_p$ and 
excess chemical potentials $\mu_c^{({\rm exc})}$, $\mu_p^{({\rm exc})}$.
One can compute the center-of-mass distribution functions
$g_{\alpha\beta,FM}(b)$ in the full-monomer system (here $\alpha$ and $\beta$
are indices that may refer to 
the polymers and to the colloids) and then
determine the CG potentials by requiring the CG model to reproduce the
functions $g_{\alpha\beta,FM}(b)$. As before, the result depends 
on the ensemble one considers, hence one obtains a different set of 
potentials for the canonical [$V_{\alpha\beta,\rm can}(b;\phi_c,\phi_p)$], 
the semigrand-canonical [$V_{\alpha\beta,SG}(b;\phi_c,\mu_p^{(\rm exc)})$], and 
the grand-canonical ensemble 
[$V_{\alpha\beta,GC}(b;\mu_c^{(\rm exc)},\mu_p^{(\rm exc)})$].

A determination of the state-dependent potentials as a function of the 
volume fractions or of the chemical potentials is equivalent to 
a complete determination of the thermodynamics of the system, so that
the use of state-dependent interactions would have little predictive power.
Ref.~\onlinecite{BLH-02} suggested to consider potentials parametrized by
a single variable, the polymer volume fraction $\phi_p^{(r)}$
of a polymer reservoir in osmotic equilibrium with the mixture.
Their approach 
works as follows. In the semigrand canonical ensemble,
each state point is characterized by the polymer 
excess chemical potential $\mu_p^{({\rm exc})}$ and by the 
colloid volume fraction $\phi_c$. Instead of using $\mu_p^{({\rm exc})}$, 
one can equivalently (but note that the equivalence only holds in the 
original, full-monomer system) consider 
the polymer volume fraction $\phi_p^{(r)}$.
If an accurate expression of the 
polymer equation of state is available, $\phi_p^{(r)}$ can be determined by
inverting the relation
\begin{equation}
\beta \mu^{(exc)}_p = \int_0^{\phi_p} 
    {d\sigma\over \sigma} (K_p(\sigma,0) - 1),
\label{mu-phi}
\end{equation}
where
\begin{equation}
K_p(\phi_p,\phi_c) = 
  \left( {\partial \beta P\over \partial \rho_p} \right)_{\rho_c}.
\label{defKp}
\end{equation}
Then, the potentials of the CG model at $\phi_c$ and $\mu_p^{({\rm exc})}$ 
is defined as \cite{footnote_BLH02}
\begin{equation}
V_{\alpha\beta,SG} (b;\phi_c,\mu_p^{({\rm exc})}) = 
   V_{\alpha\beta,\rm can} (b;0,\phi_p^{(r)}),
\label{V-statedep-SG}
\end{equation}
where the right-hand side is computed at zero colloidal density.
In practice, $V_{cc,\rm can}(b;0,\phi_p^{(r)})$ is the usual 
hard-core pair potential, while 
$V_{pp,\rm can}(b;0,\phi_p^{(r)})$ is the canonical potential 
defined before in the case of the single-component polymer system.  
The colloid-polymer potential 
$V_{cp,\rm can}(b;0,\phi_p^{(r)})$ is also determined in 
the canonical ensemble at $\phi_p^{(r)}$.
One determines the polymer density profile $g_{cp,FM}(b)$ 
around a colloid as a function of $b$ and then fixes the potential 
by requiring the CG model to reproduce $g_{cp,FM}(b)$ in the 
{\em canonical ensemble}.

Choice (\ref{V-statedep-SG}) is by no means unique and indeed, 
a conceptually equivalent approximation in the
semigrand-canonical ensemble is (model SB-SG)
\begin{equation}
V_{\alpha\beta,SG} (b;\phi_c,\mu_p^{({\rm exc})}) = 
   V_{\alpha\beta,SG} (b;0,\mu_p^{({\rm exc})}),
\label{V-statedep-SG2}
\end{equation}
where the right-hand side is computed at zero colloidal density.
The potentials in the right-hand side are obtained by considering
the same target functions $g_{pp,FM}(b)$ and $g_{cp,FM}(b)$ as before,
but now the equality of the structural properties 
is obtained in the polymer grand-canonical ensemble at zero colloidal density.

\begin{figure}[t]
\begin{center}
\begin{tabular}{c}
\includegraphics[width=8cm]{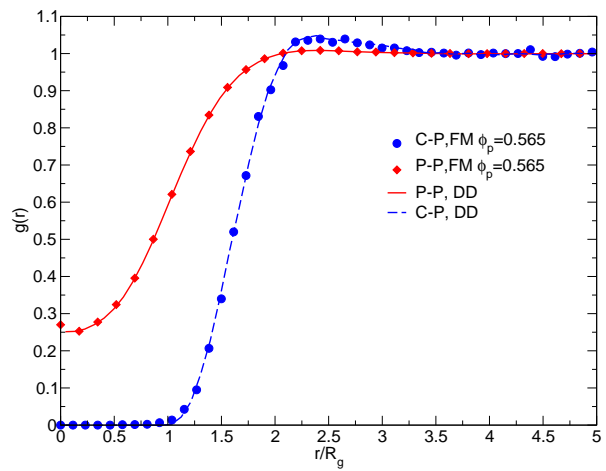}
\end{tabular}
\end{center}
\caption{Polymer-polymer and polymer-colloid distribution functions (for the 
polymers we consider its center of mass). We report 
full-monomer data and canonical-ensemble 
simulation results for the density-dependent (DD) single-blob (SB-can) 
model with 
potentials (\ref{V-statedep-can}) at $\phi_p = 0.565$ and 
zero colloidal density. Here $q=1$.
}
\label{fig:DDPOT1}
\end{figure}

If one is only interested in properties of the homogeneous phase, one 
might consider the mixture in the canonical ensemble. Again, the choice of 
the potential set is ambiguous. Here we consider two possibilities. 
First, we define (model SB-can) 
\begin{equation}
V_{\alpha\beta,\rm can} (b;\phi_c,\phi_p) = 
   V_{\alpha\beta,\rm can} (b;0,\phi_p^{(r)}),
\label{V-statedep-can}
\end{equation}
where the reservoir polymer volume fraction $\phi_p^{(r)}$ 
is defined before. Another 
possibility is simply (model SB-$\phi_p$)
\begin{equation}
V_{\alpha\beta,\rm can} (b;\phi_c,\phi_p) = 
   V_{\alpha\beta,\rm can} (b;0,\phi_p).
\label{V-statedep-can2}
\end{equation}
As a case study,
we first consider a polymer-colloid mixture with $q=1$, the value of $q$ 
where the zero-density single-blob model begins to break down.
We take $\phi_c=0.2$, $\phi_p=0.2$, and, as in Sec.~\ref{sec5}, we 
take the Domb-Joyce model with $L=600$ as reference system. 
At this state point 
the tetramer model reproduces correctly the structure and the thermodynamics 
of the mixture (see supplementary material\cite{suppl}) 
and  can be used to obtain some quantities hardly measurable with 
full-monomer simulations of long polymers. One such quantity is the polymer 
chemical potential, which can be determined by using Widom's insertion 
method. In the tetramer model we obtain $\beta \hat{\mu}_p=1.15356$, where 
\begin{equation}
  \beta\hat{\mu}_p = \log \phi_p + \beta{\mu}_p^{(exc)}.
\end{equation}
The quantity $\hat{\mu}_p$ differs from $\mu_p$ by a 
density-independent constant that depends on the detailed intramolecular 
structure, but it has the advantage that, at a given state point, it is the 
same in the full-monomer model and in the CG ones.
Using the accurate equation of state of Ref.~\onlinecite{Pelissetto-08} 
and Eq.~(\ref{mu-phi}), we obtain $\phi_p^{(r)} = 0.565$.
Once $\phi_p^{(r)}$ and $\beta\hat{\mu}_p$ for the reservoir are known,
we should compute the CG potentials. Instead of performing a 
direct numerical inversion, using the iterative Boltzmann inversion method
for instance, one can use integral-equation methods.
\cite{LBHM-00,BLHM-01,LBMH-02,BL-02}
For both ensembles, we use the hypernetted-chain (HNC) approximation which 
turns out to be quite accurate (see App.~\ref{App.C} for a discussion of 
the grand-canonical case).
Once the potentials have been 
obtained, we have verified their accuracy. For each model, we have performed 
Monte Carlo simulations in the appropriate ensemble,
computing the distribution 
functions and comparing the results with the full-monomer estimates.
As an example, in Fig.~\ref{fig:DDPOT1} we show the results obtained 
in canonical-ensemble simulations of the model with 
potentials (\ref{V-statedep-can}). Comparison with 
the full-monomer target functions shows that 
the inversion procedure is quite accurate.
The quality of the inversion can also be tested by 
computing some thermodynamic observables that are related to the 
target structural quantities through simple sum rules, and that are 
strongly influenced by the accuracy of the tails of the 
potentials. For instance, we have computed the pressure 
derivative $K_p$, Eq.~(\ref{defKp}). 
In the canonical ensemble at $\phi_p^{(r)}=0.565$ 
we obtain $K_p=2.95(5)$ for model (SB-can), 
which is in good agreement with the full-monomer 
estimate $K_p=2.94$, obtained by using the equation of state of
Ref.~\onlinecite{Pelissetto-08}. Analogously, we compute the 
polymer depletion thickness,
finding  $\delta_s/R_c = 0.55(1)$ to be 
compared with the full-monomer result $0.53(3)$. 
It is also interesting to compute the same quantities in the 
tetramer model: we have  $K_p=2.90(4)$  and $0.54(1)$, respectively,
again in good agreement with the full-monomer predictions.
Note that our potentials differ from those reported in 
Ref.~\onlinecite{BL-02}, since here we consider polymers in the 
scaling limit, while the potentials of Ref.~\onlinecite{BL-02} are 
appropriate for self-avoiding walks with 500 monomers.

The nonequivalence of the two ensembles for density-dependent
models implies that not all quantities
are reproduced by the CG model, even at $\phi_c = 0$.\cite{DPP-13-state-dep}
For instance, let us consider the pure polymer model in the 
canonical ensemble at $\phi_p^{(r)} = 0.565$ and let us compute the 
chemical potential by Widom's method. The CG model gives
$\beta \hat{\mu}_p = 1.2530(2)$, which is different from the 
the value of the chemical potential at the reference point. Viceversa, 
if we consider the grand-canonical ensemble  at $\hat{\mu}_p = 1.15356$ 
we obtain a volume fraction $\phi_p^{(r)} = 0.546$, which differs
by 3.5\% from the reference value.

\begin{table}[tbp!]
\caption{Comparison of single-blob results for state point 
$q=1,\phi_c=0.2,\phi_p=0.2$. We report canonical results for the 
full-monomer (FM) model, the zero-density single-blob (SB) model, 
the canonical density-dependent single-blob (SB-can) model 
[potentials (\ref{V-statedep-can})], and 
the density-dependent single-blob (SB-$\phi_p$) model at fixed $\phi_p$
[potentials (\ref{V-statedep-can2})]. 
Results labelled SB-SG are obtained in semigrand-canonical simulations 
using potentials (\ref{V-statedep-SG2}). The input quantities are reported in 
boldface. The FM value for $\beta \hat{\mu}_p$ (in brackets)
has been determined by using 
the tetramer model. See App.~\ref{Appendix-KB} for the definitions of the 
thermodynamic quantities.
}
\label{tab:DD-therm}
\begin{center}
\begin{tabular}{ccccccc}
\hline\hline
 & FM & SB & SB-can & SB-SG & SB-$\phi_p$  \\
\hline
\squeezetable
$\beta \hat{\mu}_p $
      & [1.15356] &  0.9970(6) & 1.2898(4) &  {\bf 1.15356} & 1.0784(3)   \\
$\phi_p$ & {\bf 0.2} & {\bf 0.2} & {\bf 0.2} & 0.18360(3) & {\bf 0.2} \\
$S_{pp,0}$ & 1.48(10) & 1.44(1)& 1.86(2) & 1.85(2) & 1.51(1)\\
$S_{cp,0}$ & $-0.66(5)$& $-0.619(8)$& $-0.86(1)$&$-0.85(1)$&$-0.658(9)$\\
$S_{cc,0}$ & 0.44(2)&  0.428(5)& 0.562(7)  &  0.553(8) & 0.449(5)\\
$K_p$ 
& 5.01(6) &  4.42(3) & 4.76(1) & 4.70(1) & 4.54(1)\\
$K_c$ 
& 9.72(5) &  8.73(6)&  9.09(4) & 8.72(4)  & 8.88(3)\\
$\frac{\beta R_c^3}{\chi_T}$ & 0.703(5)&  0.63(4)& 0.661(2) & 0.622(2) & 0.641(2)\\
$1/g''$ & 0.41(3)&  0.389(4)& 0.518(6) &  0.519(6) & 0.409(4) \\
\hline
\hline
\end{tabular}
\end{center}
\end{table}

Once we have determined the density-dependent potentials, we can use them
to compute thermodynamic properties of the mixture.
First, we compare two 
different choices of potentials 
that look most natural: 
a) we perform semigrand-canonical 
simulations at $\hat{\mu}_p = 1.15356$ and $\phi_c = 0.2$ using the 
CG model with potentials (\ref{V-statedep-SG2});
b) we perform canonical 
simulations at $\phi_p = 0.2$ and $\phi_c = 0.2$ using the CG model 
with potentials (\ref{V-statedep-can}).
We compare the 
results of these simulations with canonical 
full-monomer and zero-density single-blob estimates. 
Results are reported in Table~\ref{tab:DD-therm}. 
For the zero-momentum structure factors, the estimates 
obtained in state-dependent models are significantly worse 
than the results of the zero-density single-blob model. Also the 
second-order derivative of the Gibbs free energy $g''$ (see 
App.~\ref{Appendix-KB}) is determined more accurately by the standard 
single-blob model than by the state-dependent ones. Since 
this quantity is an order parameter for the critical point, 
it is not clear if phase behavior is determined more accurately
by using single blob models defined at zero density or by 
models with state-dependent interactions. On the other hand,
for some other observables, like the pressure derivatives $K_p$ and $K_c$,
the state-dependent models give estimates that are closer to the 
full-monomer results than those of the single-blob model defined at zero density.

As a final remark, let us consider the canonical CG 
model at $\phi_p = 0.2$ and $\phi_c = 0.2$ with potentials
(\ref{V-statedep-can2}).
Since the polymer density at which the potentials are computed is small, results 
are not too different from those of the zero-density single-blob model.
Discrepancies with the full-monomer result are however smaller. 
Note, however, that this approach cannot be used to investigate 
phase separation, since the coexisting phases would be associated with 
different pair potentials. This is not the case of potentials 
(\ref{V-statedep-SG}) and (\ref{V-statedep-SG2}), which could both 
be employed to estimate the fluid-fluid binodals.

The same study has been performed for $q=0.5$ at the thermodynamic state 
$\phi_c=0.2,\phi_p=0.1$, at which the tetramer model predicts a 
chemical potential $\hat{\mu}_p=-1.05322$, that corresponds to a 
reservoir volume fraction $\phi_p^{(r)}\approx 0.2$. Since $\phi_p^{(r)}$ is small,
the ensemble dependence of the potentials is tiny. Moreover, they show
only small differences with respect to their zero-density counterparts.
Results are compared in Table~\ref{tab:DD-therm-2}. For this value of 
$q$, the state-dependent results are in better agreement with 
full-monomer and tetramer results than the single-blob estimates, which are,
however, already in reasonable agreement. 

The analysis presented here shows that the use of state-dependent 
potentials does not provide a systematic improvement with respect to the 
zero-density single-blob model. 
Apparently, there is no clear advantage in using state-dependent
potentials, instead of those defined at zero polymer density.


\begin{table*}[tbp!]
\caption{Comparison of single-blob results for state point 
$q=0.5,\phi_c=0.2,\phi_p=0.1$. Symbols are defined as in
Table~\ref{tab:DD-therm}. 
}
\label{tab:DD-therm-2}
\begin{center}
\begin{tabular}{cccccc}
\hline\hline
& FM & SB & SB-can & SB-GC & $n=4$  \\
\hline
\squeezetable
\footnotesize
$\beta \hat{\mu}_p$& &  $-1.0659(1)$ & $-1.0483(1)$ &  
   ${\bf -1.0532}$& $-1.0532(1)$   \\
$\phi_p$ & {\bf 0.1} & {\bf 0.1} & {\bf 0.1} & 0.09956(1) & {\bf 0.1} \\
$S_{pp,0}$ & 1.86(7) & 1.674(5)& 1.759(5) & 1.73(2) & 1.713(8)\\
$S_{cp,0}$ & $-0.62(2)$& $-0.553(3)$& $-0.591(3)$ &$-0.577(9)$&$-0.572(4)$\\
$S_{cc,0}$ & 0.339(8)&  0.317(2)& 0.333(2)  &  0.325(5) & 0.322(2)\\
$K_p$ & 2.64(2) &  2.648(2) & 2.671(2) & 2.673(6) & 2.706(4)\\
$K_c$ & 12.62(6) &  12.41(2)&  12.51(2) & 12.54(2)  & 12.72(3)\\
$\frac{\beta R_c^3}{\chi_T}$ & 1.107(6)&  1.098(1)& 1.107(1) & 1.107(4) & 
 1.112(2)\\
 $1/g''$ & 0.182(6)&  0.1649(8)& 0.1746(8) &  0.172(2) & 0.169(1) \\
 \hline
 \hline
\end{tabular}
\end{center}
\end{table*}

\section{Conclusions} \label{sec8}

In this paper we determine a fully consistent multiblob model for mixtures of 
hard-sphere colloids and linear polymers under good-solvent conditions. 
We use the structure-based route, determining the effective potentials
at zero polymer and colloidal density. This allows us to avoid all ambiguities
related to the use of state-dependent interactions.
\cite{Louis-02,DPP-13-state-dep} 
Moreover, at zero density it is easy to compute 
properties in the scaling limit, which are then used as target functions 
to construct the CG model. As a consequence, the resulting 
CG models allow us to determine 
thermodynamic and structural properties directly in the scaling limit with 
a limited computation effort. Hence, no extrapolations in the polymer length 
are needed before comparing with the results of experiments 
with high molecular-weight polymers.  As in our 
previous work, \cite{DPP-12-Soft,DPP-12-JCP} we start by representing polymers 
with a tetramer chain of four blobs and parametrizing polymer-colloid interactions
with blob-colloid pair potentials. We show that such a model is quite accurate 
in the homogeneous phase for $q\lesssim 1$. It reproduces both the intramolecular 
and intermolecular structure on scales $r\gtrsim \hat{r}_g$, where 
$\hat{r}_g$ is the zero-density blob radius of gyration. Also thermodynamics is 
well reproduced. For $q=2$, we observe small differences between tetramer and 
full-monomer estimates, which increase with colloid and polymer densities. 
To investigate the behavior of the mixture for larger values of $q$, higher-resolution 
models are needed. We show that a simple transferability assumption of the 
blob-colloid potentials makes the model fully transferable 
with the number $n$ of blobs. Indeed, the blob-colloid pair potentials 
for $n > 4$ can be obtained from the 
tetramer ones by performing simple length rescalings. The basic assumption, which is 
confirmed by the numerical results, is that potentials are resolution independent,
if the blob radius of gyration $\hat{r}_g$ is taken as reference length scale and if the 
ratio $q_b = \hat{r}_g/R_c$ is assumed as reference polymer-to-colloid size ratio.
We explicitly consider the decamer model with $n=10$ blobs. We find it to be accurate 
for $q\lesssim 2$ in the homogeneous phase, below the  demixing binodal. 

We also discuss in detail single-blob models with state-dependent potentials. 
We consider several variants---as discussed in Ref.~\onlinecite{DPP-13-state-dep}, 
state-dependent potentials depend both on the thermodynamic state and 
on the ensemble considered. 
If potentials are independent of the colloidal density, 
as is the case for the 
models discussed in Refs.~\onlinecite{BLH-02,BL-02,FBD-08}, 
the predictions of these models 
are significantly less accurate than those of the multiblob model.

Our model can be readily used to study the phase diagram of polymer-colloid
solutions under good-solvent conditions. This study is technically difficult
at the full-monomer level. Simulation studies are limited to quite 
short chains,\cite{CVPR-06,MLP-12,MIP-13} so that results are affected 
by large scaling corrections. As a consequence,
extrapolations must be performed,\cite{DPP-14-GFVT} adding a considerable 
amount of uncertainty on the final results, before comparing simulation
results 
with experimental data on high-molecular-weight polymer solutions. 

Along the lines of our previous work on polymer solutions,
\cite{DPP-13-thermal} our strategy can be extended to investigate 
mixtures in the thermal crossover region (within the GFVT approximation,
this issue has already been discussed in Ref.~\onlinecite{DPP-14-GFVT}).
The same strategy could also be extended to different systems with a 
characteristic mesoscopic length scale, for instance,
to stretched chains and networks, 
polymers of different architecture, tethered chains,
and solutions with colloids of nonspherical shape. 
Also in this case a multiblob approach should be quantitatively 
accurate, as long as the blob size is comparable or smaller than all 
characteristic length scales of the system. Transferability with the 
number of blobs should also work in general, as it is based 
on the self-similar structure of the polymers.

\appendix 
\section{Structure factors and thermodynamic properties}
\label{Appendix-KB}

In this appendix we collect some formulae that allow one
to compute thermodynamic properties from structural estimates.
For the sake of generality, let us consider a binary mixture of two
types of molecules, which have $L_1$ and $L_2$ atoms each. If there are 
$N_\alpha$ molecules of type $\alpha$ in a volume $V$, the 
structure factor $S_{\alpha\beta}({\bf k})$ is defined as 
\begin{equation}
S_{\alpha\beta}({\bf k}) = {1\over L_\alpha L_\beta \sqrt{N_\alpha N_\beta} }
       \left\langle \sum_{ij,AB}
     e^{i{\bf k}\cdot ({\bf r}_{iA}^{(\alpha)} - 
                       {\bf r}_{jB}^{(\beta)})} \right\rangle,
\end{equation}
where ${\bf r}_{iA}^{(\alpha)}$ is the position of atom $i$ belonging to
molecule $A$ of type $\alpha$. Analogously, we define the pair 
correlation function
\begin{eqnarray}
 g_{\alpha\beta}({\bf r}_1 - {\bf r}_2) &= & 
 {1\over L_\alpha L_\beta \rho_\alpha \rho_\beta } \times
\\
 &&  \left\langle {\sum_{AB}}' \sum_{ij}
     \delta({\bf r}_1  - {\bf r}_{iA}^{(\alpha)} )
     \delta({\bf r}_2  - {\bf r}_{jB}^{(\beta) } )
                       \right\rangle,
\nonumber 
\end{eqnarray}
where $\rho_\alpha = N_\alpha/V$ is the density of the $\alpha$ 
molecules and the prime in the summation over $A$ and $B$ indicates that 
terms with $A=B$ should not be considered if $\alpha=\beta$. 
Then, it is easy to show that 
\begin{eqnarray}
&& S_{\alpha\beta}({\bf k}) = 
   \sqrt{N_\alpha N_\beta} \delta_{{\bf k},{\bf 0}} + 
   \delta_{\alpha\beta} F_\alpha({\bf k}) 
\nonumber \\
&& \qquad + 
   \sqrt{\rho_\alpha \rho_\beta} 
   \int (g_{\alpha\beta}({\bf r}) - 1) e^{i{\bf k}\cdot {\bf r}} d{\bf r},
\end{eqnarray}
where 
\begin{equation}
 F_{\alpha}({\bf k}) = {1\over L_\alpha^2 N_\alpha}
       \left\langle \sum_{ij,A}
     e^{i{\bf k}\cdot ({\bf r}_{iA}^{(\alpha)} - 
                       {\bf r}_{jA}^{(\alpha)})} \right\rangle
\end{equation}
is the form factor of the $\alpha$ molecules. If we introduce the 
Kirkwood-Buff integrals \cite{KB-51}
\begin{eqnarray}
G_{\alpha\beta} = \int (g_{\alpha\beta}({\bf r}) - 1) d{\bf r},
\end{eqnarray}
we obtain 
\begin{equation}
S_{\alpha\beta,0} = \lim_{k\to 0} S_{\alpha\beta}({\bf k}) = 
1 + \sqrt{\rho_\alpha\rho_\beta} G_{\alpha\beta}.
\end{equation}
The quantities $S_{\alpha\beta,0}$ are related to thermodynamics 
by fluctuation theorems. In the grand-canonical ensemble we have 
\cite{HMD-06,BenNaim} 
\begin{equation}
{1\over V} \left( \langle N_\alpha N_\beta \rangle -
 \langle N_\alpha \rangle \langle N_\beta \rangle \right) = 
 \sqrt{\rho_\alpha\rho_\beta} S_{\alpha\beta,0}.
\label{fluctuation-theorem}
\end{equation}
Using this result we can express several thermodynamic quantities in terms 
of $S_{\alpha\beta,0}$. We define 
\begin{equation}
|S_0| = S_{11,0} S_{22,0} - S_{12,0}^2. 
\end{equation}
Then, the density derivatives of the pressure $P$
in the canonical ensemble can be written as 
\begin{equation}
K_\alpha \equiv 
   \left( {\partial\beta P\over \partial \rho_\alpha} \right)_{\rho_\beta} = 
    |S_0|^{-1} (S_{\beta\beta,0} - \sqrt{\rho_\beta/\rho_\alpha} S_{\alpha\beta,0}).
\end{equation}
If we now consider the isobaric ($P$, $N_1$, $N_2$) ensemble and write the 
Gibbs free energy as $\beta G(P,N_1,N_2) = (N_1 + N_2) g(P,x_1)$, $x_1 = N_1/(N_1 + N_2)$,
we obtain
\begin{eqnarray}
{1\over g''} &=& \left({\partial^2 g\over \partial x_1^2} \right)^{-1}_P 
\\
  &=& x_1 x_2 \left( x_1 S_{22,0} + x_2 S_{11,0} - 2 \sqrt{x_1 x_2} S_{12,0} \right),
\nonumber 
\end{eqnarray}
where $x_2 = 1 - x_1$. As for the isothermal compressibility
\begin{equation}
\chi_T = - {1\over V} \left( {\partial V\over \partial P} \right)_{N_1,N_2},
\end{equation}
we obtain
\begin{equation}
{\beta\over \chi_T} = |S_0|^{-1} 
  (\rho_1 S_{22,0} + \rho_2 S_{11,0} - 2 \sqrt{\rho_1 \rho_2} S_{12,0} ).
\end{equation}
In order to use these expressions, we must determine the structure factors in the 
limit $k\to0$. We use here the method discussed in 
Refs.~\onlinecite{Pelissetto-08,DPP-13-depletion}. We consider a cubic box 
of size $V = M^3$ and determine $S_{\alpha\beta}({\bf k})$ for the smallest values 
available in a cubic box. We choose ${\bf k}_a = (k_a,0,0)$ and 
$k_1 = 2\pi/M$, $k_2 = 2 k_1$, $k_3 = 3 k_1$, $k_4 = 4 k_1$. 
Then, we consider the approximants
\begin{eqnarray}
S_{\alpha\beta}^{(1)} &=& {4\over 3}S_{\alpha\beta}({\bf k}_1) -
                 {1\over 3}S_{\alpha\beta}({\bf k}_2),
\label{S-approximants} \\
S_{\alpha\beta}^{(2)} &=& {3\over 2} S_{\alpha\beta}({\bf k}_1) -
                 {3\over 5} S_{\alpha\beta}({\bf k}_2) +
                 {1\over 10} S_{\alpha\beta}({\bf k}_3),
\nonumber \\
S_{\alpha\beta}^{(3)} &=& {8\over 5}S_{\alpha\beta}({\bf k}_1) -
                 {4\over 5} S_{\alpha\beta}({\bf k}_2)
\nonumber \\
   && + {8\over 35}S_{\alpha\beta}({\bf k}_3) -
                 {1\over 35}S_{\alpha\beta}({\bf k}_4).
\nonumber
\end{eqnarray}
Since $k\approx 1/M$, it is easy to show that
$S_{\alpha\beta}^{(n)} = S_{\alpha\beta,0} + O(M^{-2n-2})$.
Note that we do not consider the volume corrections (of order $1/V = M^{-3}$
see, e.g., Ref.~\onlinecite{LP-61}), which affect
$S_{\alpha\beta}({\bf k})$ at fixed $k$. For the typical volumes we consider,
such corrections are negligible (see Ref.~\onlinecite{DPP-13-thermal}
for the analogous discussion
concerning the polymer-polymer distribution function).
For the values of $\phi_c$ and $\phi_p$ we investigate and for our typical
volumes, we observe some differences between $S_{\alpha\beta}^{(1)}$ and
$S_{\alpha\beta}^{(2)}$, while 
$S_{\alpha\beta}^{(2)} \approx S_{\alpha\beta}^{(3)}$ within errors.
Hence, we take approximant $S_{\alpha\beta}^{(2)}$ as our estimate of 
$S_{\alpha\beta,0}$.

In the GFVT we have direct access to the thermodynamic properties. The 
GFVT estimates of the zero-momentum factors $S_{\alpha\beta,0}$ are obtained 
by using the grand-canonical relation 
\begin{equation}
\left( {\partial \rho_\alpha\over \partial \beta\mu_\beta} 
  \right)_{GC} = \sqrt{\rho_\alpha\rho_\beta} S_{\alpha\beta,0},
\end{equation}
which is a direct consequence of Eq.~(\ref{fluctuation-theorem}).

\section{The radius of gyration of the polymer and of the blobs} \label{App.B}

\begin{table*}[tbp!]
\caption{Blob radius of gyration $r_g(n)$ for the CGR 
of the polymer in terms of $n$ blobs and radius of gyration $R_g$ of the 
polymer as a function of $q$, $\phi_c$, and $\phi_p$. All quantities
are expressed in terms of the zero-density radius of gyration 
$\hat{R}_g$ of the polymer. The effective volume fraction $\phi_{p,\rm app}$
is defined as the volume fraction of the 
pure polymeric system for which one has the same 
value of the ratio $R_g/\hat{R}_g$, i.e. $R_g/\hat{R}_g = f_g(\phi_{p,\rm app})$,
where $f_g(\phi_p)$ is defined in Eq.~(\protect\ref{eq:fg}).
}
\label{tab:rgblob}
\begin{center}
\begin{tabular}{ccccccccc}
\hline\hline
$q$ & $\phi_c$ & $\phi_p$ & 
  $r_g(4)/\hat{R}_g$ & $r_g(10)/\hat{R}_g$  & $r_g(20)/\hat{R}_g$  & 
  $r_g(30)/\hat{R}_g$ & $R_g/\hat{R}_g $ & $\phi_{p,\rm app}$ \\
\hline
  & 0.0 & 0.0 & 0.4518  & 0.2654  & 0.1771 & 0.1397 & 1 & 0 \\
\hline
0.5& 0.2& 0.1 & 0.4484  & 0.2648  & 0.1773 & 0.1401 & 0.9766 & 0.624 \\
\hline
1 & 0.1 & 0.6 & 0.4439  & 0.2635  & 0.1767 & 0.1398 & 0.9578 & 1.184   \\
  & 0.1 & 0.8 & 0.4422  & 0.2629  & 0.1765 & 0.1396 & 0.9503 & 1.431   \\
  & 0.1 & 1.0 & 0.4405  & 0.2624  & 0.1763 & 0.1395 & 0.9437 & 1.666  \\
  & 0.2 & 0.2 & 0.4431  & 0.2634  & 0.1767 & 0.1398 & 0.9508 & 1.416   \\
  & 0.2 & 0.4 & 0.4411  & 0.2627  & 0.1765 & 0.1396 & 0.9424 & 1.712 \\
  & 0.3 & 0.2 & 0.4376  & 0.2618  & 0.1762 & 0.1395 & 0.9219 & 2.554 \\
\hline
2 & 0.1 & 0.6 & 0.4400  & 0.2622  & 0.1762 & 0.1395 & 0.9432 & 1.684 \\
  & 0.1 & 1.0 & 0.4367  & 0.2611  & 0.1758 & 0.1392 & 0.9301 & 2.198 \\
  & 0.2 & 0.4 & 0.4326  & 0.2599  & 0.1754 & 0.1390 & 0.9102 & 3.125 \\
  & 0.2 & 0.8 & 0.4291  & 0.2586  & 0.1749 & 0.1387 & 0.8974 & 3.837 \\
  & 0.3 & 0.2 & 0.4229  & 0.2568  & 0.1742 & 0.1384 & 0.8671 & 6.023 \\
\hline
4 & 0.3 & 0.2 & 0.4018  & 0.2480  & 0.1702 & 0.1360 & 0.8121 & 12.67 \\
\hline
\hline
\end{tabular}
\end{center}
\end{table*}

In this Appendix we discuss how the sizes of the polymers and of the blobs
change in the homogeneous phase as $\phi_c$ and $\phi_p$ vary. In 
Table~\ref{tab:rgblob} we report full-monomer
results for $L=600$ Domb-Joyce chains---they are not asymptotic, but we expect
differences to be relatively small. Let us first consider the ratio
$R_g(\phi_c,\phi_p)/\hat{R}_g$, where $\hat{R}_g$ is the zero-density
radius of gyration. For $\phi_c = 0$ the ratio 
$R_g(0,\phi_p)/\hat{R}_g = f_g(\phi_p)$ was computed in 
Refs.~\onlinecite{CMP-06-size,Pelissetto-08}, obtaining the interpolation
formula 
\begin{equation}
f_g(\phi_p) = {(1 + 0.33272 \phi_p)^{0.0575} \over 
  (1 + 0.98663 \phi_p + 0.49944 \phi_p^2 + 0.049597 \phi_p^3)^{0.0575} }.
\label{eq:fg}
\end{equation}
The size decreases as $\phi_p$ increases, but quite slowly:
$f_g(\phi_p) \approx \phi_p^{-0.11}$ for large $\phi_p$. For $\phi_c\not=0$,
the data show that the size of the polymers depends crucially on $q$ 
and that, for the same volume fractions $\phi_c$ and $\phi_p$, 
polymers become more compact as $q$ increases. For instance, for 
$\phi_c = 0.3$ and $\phi_p = 0.2$, we have 
$R_g(\phi_c,\phi_p)/\hat{R}_g = 0.92, 0.87, 0.81$ for $q=1,2$ and 4,
respectively. This phenomenon, which has already been noted 
in Ref.~\onlinecite{SBHHS-14} for $\phi_p = 0$, is connected to the sharp
decrease of the free-volume factor as $q$ increases.
When $q$ gets larger
at fixed $\phi_c$ and $\phi_p$, the available space for the 
insertion of the polymer decreases, hence polymers become more compact.

In Table \ref{tab:rgblob} we also report the average 
radius of gyration $r_g(n)$
of the blobs for different CGRs of the polymers. The $n$ dependence 
of the zero-density quantity $\hat{r}_g(n)$ was discussed in 
Ref.~\onlinecite{DPP-12-Soft}, where it was shown that for all 
$n\ge 4$ one can write
\begin{equation}
   {\hat{r}_{g}(n)\over \hat{R}_{g}} = 
    k n^{-\nu} \qquad k = 1.03 - 0.04/n.
\end{equation}
Here we discuss its behavior in the homogeneous phase with the purpose of 
verifying one of the basic assumptions of the CG approach. The $n$-blob 
CG model is predictive as long as the 
structure of the blobs does not play any role in the determination of the 
large-scale properties of the system.  This implies that the CG model 
provides a good approximation at ($\phi_c, \phi_p$), 
if $r_g(\phi_c,\phi_p,n) \approx \hat{r}_g(n)$.
Data shown in Table \ref{tab:rgblob} support this approximate 
equality. In all cases, differences decrease with $n$---the larger $n$ is, 
the more accurate the CG description is---and increase as $q$ gets larger---
the accuracy of the $n$ blob model worsens as $q$ increases.

\section{Grand-canonical single-blob models and integral equations} \label{App.C}

The state-dependent single-blob potentials can be accurately determined 
by using integral-equation methods.\cite{HMD-06} 
Canonical-ensemble potentials (models SB-can and SB-$\phi_p$) 
are derived as in Refs.~\onlinecite{LBHM-00,BL-02}.
The grand-canonical potentials $V_{\alpha\beta,SG}(b;0,\mu^{(\rm exc)}_p)$
are determined analogously, using the HNC relation\cite{Morita-60,Attard-91}
between chemical potential and density $\rho_p$,
\begin{equation}
\mu^{(\rm exc)}_p = {\rho_p\over2}
   \int d^3{\bf r} \left[ h_{pp}({\bf r})^2 - h_{pp}({\bf r}) c_{pp}({\bf r}) -
    2 c_{pp}({\bf r}) \right].
\label{zHNC}
\end{equation}
Here $h_{pp}({\bf b}) = g_{pp,FM}({\bf b};\mu^{(\rm exc)}_p) - 1$ 
and the direct correlation function
$c_{pp}({\bf b})$ is defined by the Ornstein-Zernike relation \cite{HMD-06}
\begin{equation}
h_{pp}({\bf b}) = c_{pp}({\bf b}) + 
  \rho_p \int d^3{\bf s}\,
  c_{pp}({\bf s}) h_{pp}({\bf b} - {\bf s}).
\label{OZ}
\end{equation}
Solving simultaneously Eqs.~(\ref{zHNC}) and
(\ref{OZ}) we obtain $\rho_p$ and $c_{pp}({\bf b})$.
The polymer-polymer potential follows
from the HNC closure relation:
\begin{equation}
\beta V_{pp,SG}({\bf b};\mu^{(\rm exc)}_p) = h_{pp}({\bf b}) - 
     c_{pp}({\bf b}) - \ln g_{pp,FM}({\bf b};\mu^{(\rm exc)}_p).
\label{VFCG-HNC}
\end{equation}
The polymer-colloid potential is determined as 
\cite{BL-02}
\begin{eqnarray}
&& \beta V_{cp,SG}({\bf b};\mu^{(\rm exc)}_p) 
=-\log(h_{cp}({\bf b})+1)+
\nonumber \\
&& \qquad  \rho_p \int d^3{\bf s}\,
  c_{pp}({\bf s}) h_{cp}({\bf b} - {\bf s}),
\end{eqnarray}
which is obtained by using the two-component 
Ornstein-Zernike relation\cite{HMD-06} in the limit $\rho_c \to 0$
and the HNC closure relation for the polymer-colloid
potential. Here $h_{cp}({\bf b}) = g_{cp,FM}({\bf b},\mu_p^{(exc)})-1$.

\clearpage

\section{Supplementary material: Details on the full-monomer simulations}
\label{suppl.1}

\subsection{Model and scaling corrections}
\label{suppl.1.A}

In order to obtain full-monomer estimates,
we consider the three-dimensional lattice Domb-Joyce (DJ)
model.\cite{DJ-72-suppl}
In this model the polymer solution is mapped onto $N$ chains of $L$ monomers
each
on a cubic lattice of linear size $M$ with periodic boundary
conditions. Each polymer chain is modelled by a random walk
$\{{\mathbf r}_{1A},\ldots,{\mathbf r}_{LA}\}$ with
$|{\mathbf r}_{iA}-{\mathbf r}_{i+1,A}|=1$ (we take the
lattice
spacing as unit of length) and
$1\le A \le N$. The Hamiltonian is given by
\begin{eqnarray}
H &=& \sum_{A=1}^N \sum_{1\le i < j \le L}
  \delta({\mathbf r}_{iA},{\mathbf r}_{jA})
\nonumber \\
&+ &
  \sum_{1\le A < B \le N} \sum_{i=1}^L \sum_{j=1}^L
   \delta({\mathbf r}_{iA},{\mathbf r}_{jB}),
\end{eqnarray}
where $\delta({\mathbf r},{\mathbf s})$ is the Kronecker delta.
Each configuration is weighted by $e^{-w H}$, where $w > 0$ is a free
parameter that plays the role of inverse temperature.
This model is similar
to the standard lattice self-avoiding walk (SAW) model,
which is obtained in the limit  $w \to +\infty$. For finite positive $w$
intersections are possible although energetically penalized.
For any positive $w$, this model has the same scaling limit as the
SAW model\cite{DJ-72-suppl} and thus allows us to compute the
universal scaling functions that are relevant for polymer solutions
under good-solvent conditions.
In the absence of colloids, there is a significant
advantage in using Domb-Joyce chains instead of SAWs. For SAWs the leading
scaling corrections, which decay as $L^{-\Delta}$ ($\Delta = 0.528(12)$,
Ref.~\onlinecite{Clisby-10-suppl}), are particularly strong, hence the
universal, large--degree-of-polymerization limit is only observed for
quite large values of $L$. Finite-density properties are those
that are mostly affected by scaling corrections, and indeed it is
very difficult to determine universal thermodynamic properties of polymer
solutions for $\Phi\gtrsim 5$ by using lattice SAWs.\cite{Pelissetto-08-suppl}
These difficulties are overcome by using the Domb-Joyce model for a
particular value of $w$,\cite{BN-97-suppl,CMP-06-suppl,footnote-w} 
$w = 0.505838$. 
For this value
of the repulsion parameter, the leading scaling corrections have a negligible
amplitude,\cite{BN-97-suppl,CMP-06-suppl} 
so that scaling corrections decay faster,
approximately as $1/L$.  As a consequence, scaling results are obtained
by using significantly shorter chains. All full-monomer results presented in 
this paper are obtained by using the optimal model.

\begin{table*}[tbp!]
\caption{Depletion thickness $\delta_s(0)$ at zero density and 
first polymer-density correction $\delta_1$, see text for definitions, 
for $q=0.5$, 1, and 2.
We report full-monomer (FM) results for $L=600$ Domb-Joyce walks and 
in the scaling limit 
($L=\infty$), and results for the CG model
with $n=1$ and $n=4$ blobs determined 
by using the scaling-limit polymer-colloid distribution functions ($L=\infty$)
or the distribution functions appropriate for $L=600$ Domb-Joyce walks.
}
\label{tabsuppl:depletion}
\begin{center}
\begin{tabular}{ccccccccc}
\hline\hline
\squeezetable
&& \multicolumn{3}{c}{$\delta_s(0)/R_c$} & &
  \multicolumn{3}{c}{$\delta_1$} \\ 
 \cline{3-5} 
 \cline{7-9} 
$L$ &$q$ & FM & $n=1$ & $n=4$  && FM & $n=1$ & $n=4$\\ \hline
600&0.5&0.4661(3) & 0.46639(3)& 0.4666(1)&&$-$1.061(3)&$-$1.1081(9)&$-$1.064(5)\\
$\infty$ & 0.5 	& 0.474(1)   & 0.47371(2) &0.4745(1) & &
         $-$1.05(1)    & $-$1.0924(9) &$-$1.050(5)  \\
600&1.0&0.8558(4) & 0.86209(5)& 0.8639(2)& &
         $-$1.067(2)&$-$1.1387(4)&$-$1.097(3)\\
$\infty$ & 1.0	& 0.873(4) & 0.86767(5) &0.8764(3) & &
         $-$1.052(7)  &$-$1.1311(5) & $-$1.061(4) \\
600&2.0&1.5053(5) & 
      1.50573(9)& 1.5151(3)&&$-$1.0635(10)&$-$1.1491(3)&$-$1.103(3) \\	
$\infty$ & 2.0	& 1.547(2)   & 1.54243(9) &1.5487(3)  & &
         $-$1.046(4)  &$-$1.1303(3) &$-$1.063(3)\\	
\hline\hline
\end{tabular}
\end{center}
\end{table*}

\begin{table*}[tbp!]
\caption{Virial combinations for full-monomer (FM) systems, 
for the single-blob model ($n=1$), for the tetramer ($n=4$), and 
for the decamer ($n=10$, model c) model. 
For the third-virial 
combinations, we also report the simple-liquid contribution $A_{3,\#}^I$ and
the flexibility contribution $A_{3,\#}^{\rm fl}$ (see
Ref.~\onlinecite{DPP-13-depletion-suppl}, App. A, for the definitions): 
$A_{3,\#} = A_{3,\#}^I + A_{3,\#}^{\rm fl}$. The results are obtained by using 
Domb-Joyce walks with $L=600$ monomers and the corresponding CG models (i.e., 
obtained by using $L=600$ Domb-Joyce distribution functions as targets).
Results in the scaling limit are reported in
Table~I of the paper.}
\label{tabsuppl:virial-tetramer-L600}
\begin{center}
\begin{tabular}{ccccccccc}
\hline
\hline
$q$ & $n$ & $A_{2,cp}$ & $A^I_{3,cpp}$ & $A^{\rm fl}_{3,cpp}$ & $A_{3,ccp}$  
    & $A^I_{3,ccp}$ & $A^{\rm fl}_{3,ccp}$ & $A_{3,ccp}$ \\
\hline
0.5& FM & 105.60(6) & 732.6(9) & $-$18.8(4) & 714(1) & 8522(11) & $-$124(2) & 
     8400(11) \\
& 1   & 105.901(6) & 693.9(4) & 0 & 693.4(6) & 8452(2) & 0  & 8452(2) \\
& 4   & 105.76(3) & 744(2) & $-$16(1)& 729(2)& 8557(14)&$-$119(6) &8439(13)\\
\hline
1 & FM & 26.77(1) & 134.7(2) & $-$5.8(1) & 128.9(2) & 355.2(4) & $-$10.8(1) & 
    345.2(4) \\
& 1  & 26.796(2) & 116.55(6)& 0 & 116.55(6) & 331.8(2) & 0 & 331.8(2) \\
& 4  & 27.126(9) & 135.5(4) & $-$5.0(3) & 130.5(4) & 356(1) & $-$10.0(5)& 
    346(1) \\
& $10$ & 26.920(7) & 138.5(3)& $-$6.0(2) & 132.4(4) & 360.6(6) & $-$11.0(4) & 
    349.6(6) \\
\hline
2 & FM & 8.234(5) & 26.13(3) & $-$1.67(2) & 24.45(5) &  16.60(2) & $-$0.87(1) & 
    15.73(2) \\
& 1 & 8.2866(9) & 19.16(2) & 0 &  19.16(2) & 12.55(2) & 0 & 12.55(2) \\
& 4 & 8.331(3) & 25.1(1) & $-$1.2(1) & 23.9(2) & 16.2(2) & $-$0.7(5) & 15.5(2)\\
& $10$ & 8.298(3) & 26.7(1) & $-$1.6(1)& 25.0(1)& 16.85(9)& $-$0.91(4) &
    15.94(9) \\
\hline
\hline
\end{tabular}
\end{center}
\end{table*}

It is important to stress that the leading scaling corrections are 
not related to the lattice nature of the model. Generic continuum
models show the same type of scaling corrections as lattice ones,
a result that can be proved in the renormalization-group framework.
Indeed, using the mapping between polymer models and zero-component 
spin models,\cite{deGennes-79-suppl}, one can show
\cite{CPRV-98-suppl} that the leading scaling correction related to the 
cubic lattice structure scales as $L^{-\omega_{nr} \nu}$, with 
$\omega_{nr} \approx 2$, hence it is subleading with respect to the one
that scales as $L^{-\Delta}$. 

The optimal model is particularly convenient computationally, as it 
allows us to obtain scaling-limit results by considering 
chains of moderate length. At zero density, simulations with $L=600$ chains
provide results that are essentially in the scaling limit 
(relative differences are less than 1\%), without the 
need of any extrapolation.

The Domb-Joyce model can be extended, including repulsive hard spheres of 
radius $R_c$. Their centers are not constrained to belong to the 
lattice, so that the spheres
can move everywhere in continuum space. Colloids interact with 
the polymers by means of a simple hard-core potential. The interaction
potential between a monomer and a colloid is given by 
$U_m(r) = 0$ if $r > R_c$ and $U_m(r) = \infty$ if $r < R_c$.

The nice convergence properties of the Domb-Joyce model do not hold in
the presence of repulsive colloids.
Indeed, the presence of a hard surface 
gives rise to new boundary renormalization-group operators.\cite{DDE-83-suppl}
The leading one gives rise to corrections that scale as
\cite{DDE-83-suppl} $L^{-\nu}$,
where $\nu$ is the Flory exponent (an explicit test
of this prediction is presented in the supplementary material
of Ref.~\onlinecite{DPP-13-depletion-suppl}). Because of them,
estimates of colloid-polymer properties obtained by using $L=600$ 
chains are not asymptotic (at the 1\% level). 
For instance, in Table~\ref{tabsuppl:depletion}
we report the zero-density depletion thickness $\delta_s(0)$ and 
the quantity $\delta_1$  defined by the expansion\cite{DPP-13-depletion-suppl}
\begin{equation}
{\delta_s(\phi_p) \over \delta_s(0) } = 
   1 + \delta_1 \phi_p + O(\phi_p^2),
\end{equation}
at zero colloidal density. We report the full-monomer scaling results 
(they are obtained by extrapolating finite-$L$ data, as discussed below)
and the estimates obtained by using $L=600$ Domb-Joyce chains. Differences
of the order of 2-3\% are clearly present. We also report CG estimates.
Those corresponding to $L=600$ are obtained by using the model 
that reproduces the structure of $L=600$ chains, while those labelled 
``$\infty$" are obtained by using the CG model 
meant to reproduce the polymer structure in the scaling limit.
In Table~\ref{tabsuppl:virial-tetramer-L600} we report
the virial combinations defined in Sec.~II of the paper.
Here we report the results for $L=600$ chains. The corresponding 
scaling-limit results are reported in 
Table~I of the paper.
Again, differences of the order of a few percent
between $L=600$ and scaling-limit results 
are clearly visible.

In order to obtain estimates of full-monomer quantities in the scaling
limit, an extrapolation is needed.
For this purpose, we proceed as follows. To estimate a 
universal quantity $f(L)$ in the limit $L\to \infty$, we determine
$f(L)$ for $L = L_1 = 600$ and $L = L_2 = 2400$.
Then, we assume that
only the leading scaling correction is relevant, so that the 
expansion $f(L) = f^* + a/L^\nu$ is accurate for $L\gtrsim 600$. 
The scaling-limit quantity $f^*$ is then estimated as 
\begin{equation}
f^* = {L_1^{-\nu} f(L_2) - L_2^{-\nu} f(L_1) \over 
       L_1^{-\nu} - L_2^{-\nu} } .
\end{equation}

\subsection{Algorithmic details}
\label{suppl.1.B}

The Domb-Joyce model\cite{DJ-72-suppl}
is very convenient from a computational point
of view. Since interactions are soft, the Monte Carlo dynamics for
Domb-Joyce chains is quite fast. In the full-monomer simulations 
we used the algorithm described in Ref.~\onlinecite{Pelissetto-08-suppl}, 
which is very efficient for pure polymer systems, as it allows one to obtain
precise results for quite long chains ($L\lesssim 1000$) deep in the
semidilute regime.

In order to simulate the system we use several types of moves.
\begin{itemize}
\item[(i)] We consider pivot moves
\cite{Lal-69-suppl,MJHS-85-suppl,MS-88-suppl,Sokal-95b-suppl}
applied to a single polymer.  
\item[(ii)] We translate
a polymer rigidly by one lattice site. These moves are relevant for the
diffusional dynamics of the polymers. 
\item[(iii)] We consider cut-and-permute (CP) moves.
\cite{Causo-02-suppl,Pelissetto-08-suppl}
In finite-density simulations they
represent a nonlocal generalization of the usual reptation moves.
\item[(iv)] We consider standard reptation moves.
\item[(v)] We consider random translations of the colloids.
The average step size is
chosen to obtain an average acceptance of 50\%.
\end{itemize}

\begin{table}[tbp!]
\caption{Acceptance ratios $a = N_{\rm acc}/N_{\rm prop}$ 
($N_{\rm acc}$ and $N_{\rm prop}$ are the number of accepted and proposed 
moves, respectively) for pivot moves ($a_{\rm piv}$), 
translations ($a_{\rm transl}$), and cut-and-permute moves ($a_{\rm cp}$),
as a function of $q$, $\phi_c$, and $\phi_p$. 
Results for the optimal Domb-Joyce model: chains with $L=600$ monomers.
} 
\label{tabsuppl:Acceptance}
\begin{center}
\begin{tabular}{ccccccc}
\hline\hline
$q$ & $\phi_c$ & $\phi_p$ & $a_{\rm piv}$ & $a_{\rm transl}$ & $a_{cp}$ \\
\hline
0.5 & 0.2 & 0.1 & 0.54100(3) & 0.93262(2) & 0.38275(3) \\    
\hline
1   & 0.1 & 0.6 & 0.40831(4) & 0.71848(7) & 0.22016(4) \\         
    & 0.1 & 0.8 & 0.36771(5) & 0.64211(9) & 0.18438(4) \\      
    & 0.1 & 1.0 & 0.33247(6) & 0.57291(8) & 0.15753(5) \\      
    & 0.2 & 0.2 & 0.43297(4) & 0.82749(5) & 0.23908(3) \\
    & 0.2 & 0.4 & 0.38193(3) & 0.72371(5) & 0.19200(3) \\                 
    & 0.3 & 0.2 & 0.35246(6) & 0.73942(11)& 0.16445(6) \\      
\hline
2   & 0.1 & 0.6 & 0.33675(5) & 0.64094(10)& 0.15767(4) \\
    & 0.1 & 1.0 & 0.27699(4) & 0.50203(11)& 0.11924(3) \\      
    & 0.2 & 0.4 & 0.25776(5) & 0.55217(12)& 0.10510(3) \\      
    & 0.2 & 0.8 & 0.21193(6) & 0.40122(15)& 0.08308(4) \\      
    & 0.2 & 1.0 & 0.19462(5) & 0.34366(10)& 0.07555(3) \\      
    & 0.3 & 0.2 & 0.19392(4) & 0.45249(16)& 0.07274(3) \\       
\hline
4   & 0.3 & 0.2 & 0.07832(2) & 0.04566(6) & 0.02808(2) \\     
\hline \hline
\end{tabular}
\end{center}
\end{table}

Let us now discuss the efficiency of the nonlocal polymer moves, extending
the discussion of Ref.~\onlinecite{Pelissetto-08-suppl} to the 
polymer-colloid case. Results for $L = 600$ are 
reported in Table~\ref{tabsuppl:Acceptance}. For $q\le 1$ the algorithm 
is quite efficient. Indeed, the internal structure of the chains is 
quite rapidly updated by pivot and CP moves. Moreover, chains diffuse 
quite fast, both because of the rigid translations and of the CP moves. 
Of course, if $L$ is increased the acceptance rate of the pivot and 
CP moves decreases, hence the dynamics becomes slower. Note, however, 
that the change is not large. For $q=1$, $\phi_c=0.1$ and $\phi_p=1.0$
we obtain $a_{\rm piv} = 0.277$ and $a_{\rm cp} = 0.12$ for $L=2400$ 
chains. Analogously, for $\phi_c = 0.2$ and $\phi_p = 0.4$ we obtain 
$a_{\rm piv} = 0.325$ and $a_{\rm cp} = 0.153$ for the same value of $L$.
By increasing $L$, the acceptance of the translation moves 
increases, $a_{\rm transl} = 0.67$ and $0.81$ for $\phi_c=0.1$, $\phi_p=1.0$
and $\phi_c = 0.2$, $\phi_p = 0.4$. This is probably due to the fact that,
at fixed $\phi_p$, the monomer density decreases as $L$ increases. The
improvement in the acceptance, however, does not indicate a better 
performance of the algorithm. Indeed, translations move the polymer 
only by one lattice step, while the relevant diffusion length is of the 
order of the size of the polymer, hence it scales as $L^\nu$. Therefore,
even with a larger acceptance, the diffusion dynamics becomes slower.

For $q=2$ the algorithm worsens somewhat. This is probably due to two
different factors. On the one hand, polymers are more compact, as 
discussed in App.~B. On the other hand, the available 
free space decreases, hence it becomes more difficult
to insert a large piece (whose length is of the order of $L$)
of the chain into the system. These problems become more serious for $q=4$. 
For $\phi_c = 0.3$ and $\phi_p = 0.2$, the acceptance fractions are small.
Moreover, the analysis of the local acceptance fraction, 
see Ref.~\onlinecite{Pelissetto-08-suppl} for the definition, shows that 
for $q=4$
pivot or CP moves are only accepted when the pivot or the cutting 
point are close to the chain endpoints. Essentially, only local moves 
are accepted and the CP move is not much better than reptation.

\section{Supplementary material: Explicit expressions for the colloid-blob
potentials}

We wish now to define in detail the CG model we have considered 
in this paper. The tetramer CG model used here is defined 
in Ref.~\onlinecite{DPP-12-JCP-suppl} (model 4MB-2). It differs from the 
model discussed in Ref.~\onlinecite{DPP-12-Soft-suppl} because of the 
presence of an additional angular potential. The expressions
of the potentials are given in the supplementary material of 
Ref.~\onlinecite{DPP-13-thermal-suppl} (they are more accurate than
those reported in the appendix of Ref.~\onlinecite{DPP-12-JCP-suppl}).
The decamer model is obtained by starting from the 
tetramer, using the transferability assumption.
\cite{DPP-12-JCP-suppl,DPP-13-thermal-suppl}

We report here the 
colloid-blob potentials for the tetramer. We only give the 
results appropriate for polymers in the scaling limit.
For $q=0.5$ potentials are parametrized as 
\begin{eqnarray}
\beta V_{cp,i}(x,q=0.5)&=&a_0 \exp[-b_0|x-c_0|^{d_0}]
\nonumber \\ 
 &+&
    \sum_{i=1}^{3}a_i e^{-((x-c_i)/b_i)^2}.
\end{eqnarray}
This expression parametrizes the data in the range 
$2.12 < x = r/\hat{R}_g < 7.3$.
The coefficients are reported in Table~\ref{Table-0p5-coeffs}.
For $q=1$ and 2 potentials are parametrized as 
\begin{eqnarray}
\beta V_{cp,i}(x,q=1)&=& \sum_{i=0}^{3}a_i e^{-((x-c_i)/b_i)^2}.
\end{eqnarray}
This expression parametrizes the $q=1$ potentials in the range 
$1.01 < x = r/\hat{R}_g < 5.5$ and the $q=2$ potentials in the range
$0.274 < x = r/\hat{R}_g < 4.16$.
The coefficients are reported in Table~\ref{Table-1-coeffs} for $q=1$ 
and in Table~\ref{Table-2-coeffs} for $q=2$.

\begin{table}[h]
\caption{Coefficients parametrizing the colloid-polymer potentials 
for $q=0.5$. We report results for the potentials involving the 
external (Ext) blobs and the internal (Int) blobs.}
\label{Table-0p5-coeffs}
\begin{center}
\begin{tabular}{cccccc}
\hline\hline
& $i$       & 0   &       1       &       2       &       3\\
\hline
Ext & $a_i$      & 9.096478   & 2.112644      & 0.012053 &  0.002480 \\
& $b_i$      & 7.674464   & 0.0476178     & 0.744283 & 1.515897 \\
& $c_i$      & 2.1                    &   2.1     & 3.0 & 4.0  \\
& $d_i$      & 0.922204 & ---& ---& ---\\
\hline
Int & $a_i$      & 7.103179  &        $-0.0402607$ &   $-0.0039657$ & 0\\
& $b_i$      & 9.575358  &    0.536611&   0.6702560 & 0\\
& $c_i$      & 2.099997  &        2.7     & 4.0 & 0\\
& $d_i$      & 1.062409  & ---    & ---& ---\\
\hline\hline
\end{tabular}
\end{center}
\end{table}

\begin{table}[h]
\caption{Coefficients parametrizing the colloid-polymer potentials
for $q=1$. We report results for the potentials involving the
external (Ext) blobs and the internal (Int) blobs.}
\label{Table-1-coeffs}
\begin{center}
\begin{tabular}{cccccc}
\hline\hline
& $i$       & 0   &       1       &       2       &       3\\
\hline
Ext & $a_i$      & 4.369300 & 19.97256 &  0.257137 & $-$0.250900 \\
& $b_i$      &0.227528& 0.531430 &  1.204911  & 1.198152  \\
& $c_i$      & 0.899934& 0.3837873 & 2.694420  & 2.703843 \\
\hline
Int & $a_i$      & 3.900633 & 8.612210        & $-$0.016251&   $-$0.00361436\\
& $b_i$      & 0.216330 & 0.400626        &       0.320438        &
0.97795073\\
& $c_i$      & 0.899935 & 0.701953        &       1.901012        &
3.24060230\\
\hline\hline
\end{tabular}
\end{center}
\end{table}

\begin{table}[h]
\caption{Coefficients parametrizing the colloid-polymer potentials
for $q=2$. We report results for the potentials involving the
external (Ext) blobs and the internal (Int) blobs.}
\label{Table-2-coeffs}
\begin{center}
\begin{tabular}{cccccc}
\hline\hline
& $i$       & 0   &       1       &       2       &       3\\
\hline
Ext & $a_i$      &    0.538768 &      13.735607       &       0.002287403     &
0.007868818 \\
& $b_i$      &    0.1832759  &    0.481442        &       0.5431964       &
0.165933 \\
& $c_i$      &    0.5056852        &      0       & 1.962605      &
1.1330787 \\
\hline
Int & 
$a_i$      &    9.5860832       &       46.512851       &   $-$0.01164300
&       $-$0.000896600 \\
& $b_i$      &    0.3998641       &       0.1639868       &       0.253572
&       0.488688        \\
& $c_i$      &    0.1768134       &       0                       &
1.459417                &       2.994538\\
\hline\hline
\end{tabular}
\end{center}
\end{table}

As a check we compare the virial coefficients computed
by using these parametrized
expressions for the potentials with those computed by
using the numerical potentials obtained as output of the Iterative Boltzmann 
Inversion procedure. For $q=0.5$ we find a relative difference of 
1.4\%, 0.9\%, 2.0\% for $A_{2,cp}$, $A_{3,cpp}$, and $A_{3,ccp}$.
For $q=1$ the relative difference is significantly smaller: 
0.1\%, 0.6\%, and 0.1\%, respectively. For $q=2$, we obtain
1.8\%, 1.1\%, and 0.7\% for the same quantities, respectively. 
All results presented in the paper 
were obtained by using the numerical potentials.

\section{Supplementary material: Intramolecular structure}

\begin{table}[tbp!]
\caption{Radius of gyration $R_{g,b}$ of the CGR of the polymer model 
with $n=4$ and $n=10$ blobs
(CGR-4 and CGR-10, respectively).
Results appropriate for $L=600$ Domb-Joyce chains.
We also report the radius of gyration 
for the tetramer (CG-4) and the decamer (CG-10) models.
}
\label{tabsuppl:Rgb}
\begin{center}
\begin{tabular}{ccccccc}
\hline\hline
q & $\phi_c$ & $\phi_p$ & 
CGR-4 & CG-4  & CGR-10 & CG-10  \\
\hline
\squeezetable
1 & 0.1 & 0.6 & 0.84873(3) &  0.85304(3) & 0.92086(2) & 0.91702(6) \\
&   0.1 & 0.8 & 0.84121(4) &  0.84724(4) & 0.91324(2) & \\
&   0.2 & 0.2 & 0.84120(4) &  0.84132(8) & 0.91358(2) & \\
&   0.2 & 0.4 & 0.83281(2) &  0.83408(4) & 0.90507(1) & 0.89978(4) \\
&   0.3 & 0.2 & 0.81149(6) &  0.80837(6) & 0.88400(3) & 0.87651(3) \\
\hline
2 & 0.1 & 0.6 & 0.83427(5) & 0.84037(4) & 0.90603(2) & 0.90003(4) \\
&   0.1 & 1.0 & 0.82117(5) & 0.83179(5) & 0.89268(2) & 0.88632(4) \\
&   0.2 & 0.4 & 0.80087(6) & 0.80682(5) & 0.87233(3) & 0.86106(4) \\
&   0.2 & 0.8 & 0.78822(8) & 0.80191(5) & 0.85939(3) & 0.84921(4) \\
&   0.3 & 0.2 & 0.75696(7) & 0.76645(6) & 0.82816(3) & 0.81109(3) \\
\hline
\hline
\end{tabular}
\end{center}
\end{table}

In Secs.~V and VI of the text, we discussed how the CG model reproduces 
the intramolecular structure of the full-monomer system. In particular,
we focused on the intramolecular distribution function of the 
$n$-blob CGR of the polymer model, defined by
\begin{equation}
g_{{\rm intra},n}({\bf r}) = {2 \hat{R}_g^3 \over n (n-1)} 
   \sum_{i<j} 
    \left\langle \delta ({\bf r} - {\bf s}_i - {\bf s}_j)\right\rangle,
\end{equation}
where ${\bf s}_i$ are the blob positions. For large $L$, 
$g_{{\rm intra},n}({\bf r})$ is a universal function of 
${\bf b} = {\bf r}/\hat{R}_g$, where $\hat{R}_g$ is the 
zero-density radius of gyration. At zero polymer and colloid density,
$g_{{\rm intra},n}({\bf r})$ is completely consistent \cite{DPP-12-Soft-suppl}
with the tetramer and decamer distribution functions, confirming
the accuracy of the inversione procedure (tetramer case) and of the 
transferability assumption (decamer). The same good agreement is 
observed at the set of state points we have considered in Sec.~V of
the paper, which belong to the homogeneous phase and are not too close to the
binodal, see, e.g., Fig.~7 in the main paper. 
A more quantitative check can be performed by considering the 
radius of gyration of the CGR representation of the polymer, 
defined by
\begin{equation}
{R}_{g,b}^2 = {1\over 2 n^2} \sum_{i,j=1}^n ({\bf s}_i - {\bf s}_j)^2 .
\label{Rgb-def}
\end{equation}
Such a quantity is always smaller than ${R}_g$, since
\begin{equation}
{R}_g^2 = {R}_{g,b}^2 + {1\over n} \sum_{i=1}^n {r}_{g,i}^2,
\label{Rg-Rgb}
\end{equation}
where ${r}_{g,i}$ is the radius of gyration of $i$-th blob ($m$ is the 
number of monomers belonging to each blob):
\begin{equation}
{r}_{g,i}^2 = {1\over 2 m^2}
   \sum_{k,l= m(i-1)+1}^{mi}  ({\bf r}_k - {\bf r}_l)^2.
\end{equation}
The ratios $R_{g,b}^2/R_g^2$ and $r_{g,i}^2/R_g^2$ of their averages
over the polymer configurations are universal, hence independent
of the nature of the underlying polymer model as long as $L$ is large enough.
The quantity $R_{g,b}/\hat{R}_g$ can be directly compared with the 
radius of gyration of the corresponding CG model. Estimates are reported in
Ref.~\ref{tabsuppl:Rgb}. Differences are always very small, for both $q=1$ and 
$q=2$, indicating that the CG model correctly reproduces the 
intramolecular polymer structure at the coarse-grained level.

\section{Supplementary material: Tables of thermodynamic results}

In this Section we report extensive tables of thermodynamic data
for several state points in the homogeneous phase for $q=0.5$, $q=1$, and $q=2$.
We report full-monomer results (only for $q=1$ and $q=2$), GFVT results,
and estimates obtained in the CG models with different number of blobs.
Note that full-monomer results have been obtained by using 
Domb-Joyce walks with $L=600$ monomers, without performing any extrapolation,
hence results are expected to differ by a few percent from the scaling,
universal estimates. To be consistent,
the CG models we consider in this section have been
obtained by using $L=600$ Domb-Joyce distribution functions as targets.
Therefore, differences between CG and full-monomer results are only the 
result of the inaccuracy of the CG procedure. 
On the other hand, GFVT results are 
obtained by using scaling-limit results for $K_p(\phi_p)$
and for the depletion thickness, hence they should be rather compared 
with CG and full-monomer scaling-limit results. As we already mentioned
in Sec.~\ref{suppl.1.B}
however, the difference between scaling and $L=600$ results is quite small
(a few per cent). In particular, it is significantly smaller
than the observed discrepancies between GFVT and full-monomer 
predictions, which are therefore mostly 
due to the approximate nature of the theory.

\begin{table}
\caption{Details on the full-monomer simulations of $L=600$ Domb-Joyce chains:
$N_c$ and $N_p$ are the number of colloids and polymers, respectively, 
$M$ the linear size of the cubic box, $\hat{R}_g$ the zero-density radius of 
gyration, and $N_{\rm it}$ the number of iterations. Each iteration consists
in one pivot, one cut-and-permute, and one polymer translation applied 
sequentially to each polymer, $60 N_p$ reptations, and $N_p N_c/5$ 
colloid translations. More precisely, after applying the three nonlocal moves 
to a given polymer, we perform one reptation move on 60 randomly chosen 
polymers and translate $N_c/5$ randomly chosen colloids.}
\label{full-monomer-sim}
\begin{tabular}{cccccccc}
\hline\hline
$q$ & $\phi_c$  & $\phi_p$ & $N_c$  & $N_p$  & $M$ &  $M/\hat{R}_g$  & 
     $N_{\rm it}/10^3$  \\
\hline
1   &  0.1 & 0.6 &   111 &   664 &  256 &  16.7  & $250$ \\
    &  0.1 & 0.8 &   111 &   885 &  256 &  16.7  & $200$ \\
    &  0.1 & 1.0 &   111 &  1106 &  256 &  16.7  & $200$ \\
    &  0.2 & 0.2 &   221 &   221 &  256 &  16.7  & $500$ \\
    &  0.2 & 0.2 &  1769 &  1769 &  512 &  33.3  & $150$ \\
    &  0.2 & 0.4 &   221 &   442 &  256 &  16.7  & $1000$ \\
    &  0.3 & 0.2 &   332 &   221 &  256 &  16.7  & $500$ \\
\hline
2   &  0.1 & 0.6 &   885 &   664 &  256 &  16.7  &  150  \\
    &  0.1 & 1.0 &   885 &  1106 &  256 &  16.7  &  100  \\
    &  0.2 & 0.4 &  1769 &   442 &  256 &  16.7  &  250  \\
    &  0.2 & 0.8 &  1769 &   885 &  256 &  16.7  &  150  \\
    &  0.3 & 0.2 &  2654 &   221 &  256 &  16.7  &  300  \\
\hline\hline
\end{tabular}
\end{table}

Details on the full-monomer runs are reported in Table~\ref{full-monomer-sim}.
Results for $q=0.5$ are reported in Tables 
\ref{tabsuppl:t0p5_0p1_0p4},
\ref{tabsuppl:t0p5_0p3_0p1}, and
\ref{tabsuppl:t0p5_0p3_0p2}. For this value of $q$, tetramer results 
should be accurate and also GFVT should be reliable,
hence we can use the latter to estimate the boundary of the homogeneous phase. 
For $\phi_c = 0.1$, phase separation occurs for $\phi_p \gtrsim 0.5$, 
hence the state point $\phi_c = 0.1$ and $\phi_p = 0.4$ is not far 
from the binodal. In this case, GFVT gives results which 
do not differ significantly from the tetramer ones. Note also that 
GFVT is more accurate than the single-blob model. For $\phi_c = 0.3$ and 
$\phi_p = 0.1$, GFVT, single-blob and tetramer results are close. 
For $\phi_c = 0.3$ and $\phi_p = 0.2$, GFVT predicts phase separation. 
It is not clear whether phase separation also occurs in the single-blob
and/or in the tetramer model (the results we report are obtained in canonical
simulations). Hints of phase separation are provided by
the quite large value of $S_{pp,0}$.

\begin{table}[tbp!]
\caption{Estimates of several thermodynamic quantities 
for $q=0.5$, $\phi_c=0.1$, $\phi_p=0.4$. Definitions are given in 
App.~A of the paper. We report 
GFVT estimates and results obtained in CG models with 
$n=1$ and $n=4$ blobs.
}
\label{tabsuppl:t0p5_0p1_0p4}
\begin{center}
\begin{tabular}{cccc}
\hline\hline
& GFVT & $n=1$ & $n=4$ \\
\hline
$S_{pp,0}$ & 3.147 &	2.76(1) & 3.5(2) \\
$S_{cp,0}$ & $-$1.724 &	$-$1.57(2) &	$-$2.0(1) \\
$S_{cc,0}$ & 1.0781 &	1.05(2) &	1.36(8) \\
$K_p$ & 3.295 & 3.068(6) & 3.28(2) \\
$K_c$ & 30.74 & 27.0(1) &  28.9(2) \\
$\frac{\beta R_c^3}{\chi_T}$ & 3.251 & 2.99(6) & 3.20(2) \\
$1/g''$ & 0.051 & 0.0480(7) & 0.062(4) \\ 
\hline
\hline
\end{tabular}
\end{center}
\end{table}

\begin{table}[tbp!]
\caption{Estimates of several thermodynamic quantities 
for $q=0.5$, $\phi_c=0.3$, $\phi_p=0.1$. Definitions are given in 
App.~A of the paper. We report 
GFVT estimates and results obtained in CG models with 
$n=1$ and $n=4$ blobs.
}
\label{tabsuppl:t0p5_0p3_0p1}
\begin{center}
\begin{tabular}{cccc}
\hline\hline
& GFVT & $n=1$ & $n=4$ \\
\hline
$S_{pp,0}$ & 2.69 & 2.20(1) & 2.33(3) \\
$S_{cp,0}$ & $-$0.70& $-$0.55(4) & $-$0.59(1) \\
$S_{cc,0}$ & 0.24 &	0.197(1) & 0.208(3) \\
$K_p$ & 4.318 & 4.01(1) & 4.26(1) \\
$K_c$ & 24.69 & 23.30(7) & 24.6(1) \\
$\frac{\beta R_c^3}{\chi_T}$ & 2.59 & 2.43(6) & 2.58(1) \\
$1/g''$ & 0.30 & 0.244(2) & 0.261(4) \\
\hline
\hline
\end{tabular}
\end{center}
\end{table}

\begin{table}[tbp!]
\caption{Estimates of several thermodynamic quantities 
for $q=0.5$, $\phi_c=0.3$, $\phi_p=0.2$. Definitions are given in 
App.~A of the paper. We report 
results obtained in CG models with 
$n=1$ and $n=4$ blobs.
We do not report the GFVT estimates, since this point belongs 
to the region in which GFVT predicts phase separation. 
}
\label{tabsuppl:t0p5_0p3_0p2}
\begin{center}
\begin{tabular}{ccc}
\hline\hline
& $n=1$ & $n=4$ \\
\hline
$S_{pp,0}$ & 4.78(6) & 6.19(3) \\
$S_{cp,0}$ & $-$1.32(2) &	$-$1.74(1) \\ 
$S_{cc,0}$ & 0.408(6) &	0.43(3)  \\
$K_p$ & 4.69(1) & 5.20(3) \\
$K_c$ & 37.4(1) & 41.5(2) \\
$\frac{\beta R_c^3}{\chi_T}$ & 4.47(1) & 4.96(2) \\
$1/g''$ & 0.274(4) & 0.36(2) \\
\hline
\hline
\end{tabular}
\end{center}
\end{table}

Thermodynamic results for $q=1$ and $\phi_c = 0.1$ are reported in Tables
\ref{tabsuppl:t1_0p1_0p6}, \ref{tabsuppl:t1_0p1_0p8}, and 
\ref{tabsuppl:t1_0p1_1p0}. For this value of 
$\phi_p$, the tetramer estimates are always consistent with the full-monomer 
ones, confirming its accuracy for $q=1$. 
Single-blob results differ instead significantly. Note, in particular,
that $|S_{\alpha\beta,0}|$ is relatively small and 
does not change significantly as $\phi_p$ 
is increased from 0.8 to 1.0, a behavior which indicates that the 
single-blob model undergoes phase 
separation, assuming it occurs, only for $\phi_p \gg 1$. 
GFVT predicts phase separation 
for $\phi_p \approx 0.72$, which is consistent with the somewhat large 
value of $S_{pp,0}$ for $\phi_p = 0.6$. For this value of 
$\phi_p$, the GFVT estimates of $|S_{\alpha\beta,0}|$ and of $1/g''$ are not 
consistent with the full-monomer ones. Differences are instead significantly
smaller in the case of $K_p$, $K_c$, and $\beta R_c^3/\chi_T$. 

The results for $\phi_c = 0.2$, reported in Tables
\ref{tabsuppl:t1_0p2_0p2} and \ref{tabsuppl:t1_0p2_0p4}, 
confirm what observed for $\phi_c = 0.1$. The tetramer model accurately 
reproduces the full-monomer results. GFVT predicts a critical point 
for $\phi_{c,\rm crit} = 0.178$ and $\phi_{p,\rm crit} = 0.474$ (this 
explains the large estimates of the zero-momentum structure factors in 
Table \ref{tabsuppl:t1_0p2_0p4}). Results for $\phi_c = 0.3$ are 
reported in Table \ref{tabsuppl:t1_0p3_0p2}. While tetramer and decamer 
results are consistent with the full-monomer ones, GFVT and single-blob 
results differ somewhat. As observed before, GFVT appears to be more 
accurate than the single-blob model.

Results for $q=2$ are reported in Tables
\ref{tabsuppl:t2_0p1_0p6}, \ref{tabsuppl:t2_0p1_1p0}, \ref{tabsuppl:t2_0p2_0p4},
\ref{tabsuppl:t2_0p2_0p8}, and \ref{tabsuppl:t2_0p3_0p2}. 
Comparing the full-monomer and the tetramer results, we find that 
CG model is reasonably accurate except for 
$\phi_c = 0.2$ and $\phi_p = 0.8$. This state point, however, is not too far
from the full-monomer critical point, $\phi_{c,\rm crit} \approx 0.19$, 
$\phi_{p,\rm crit} \approx 1.18$, which was estimated in 
Ref.~\onlinecite{DPP-14-GFVT-suppl}. Therefore, we conclude that
the tetramer is reasonably accurate in the homogeneous phase also for $q=2$,
except close to the critical point. No differences are observed instead 
between decamer and full-monomenr results. 
As expected, GFVT and single-blob results 
differ significantly from the correct ones.

\begin{table}[tbp!]
\caption{Estimates of several thermodynamic quantities 
for $q=1$, $\phi_c=0.1$, $\phi_p=0.6$. Definitions are given in 
App.~A of the paper. We report full-monomer (FM), 
GFVT estimates, and results obtained in CG models with 
$n=1$, $n=4$, and $n=10$ blobs.
}
\label{tabsuppl:t1_0p1_0p6}
\begin{center}
\begin{tabular}{cccccc}
\hline\hline
& FM & GFVT & $n=1$ & $n=4$ & $n=10$ \\
\hline
$S_{pp,0}$ & 1.46(4) & 2.352 & 1.25(2)& 1.39(4) & 1.41(4)\\ 
$S_{cp,0}$ & $-$1.20(4)& $-$2.030 &$-$1.00(2)& $-$1.13(4) & $-$1.16(4)\\
$S_{cc,0}$ & 1.22(4)& 1.985 & 1.06(2)& 1.15(4) & 1.19(4)\\
$ K_p$ & 5.11(6) & 5.14  & 4.32(2) &4.98(3) & 5.13(2)\\
$ K_c$ & 13.1(2) &13.39 & 10.85(7)& 12.85(8) & 13.15(6)\\
$\frac{\beta R_c^3}{\chi_T}$ & 1.05(1)& 1.0564 & 0.887(5)& 1.021(5)& 1.049(4)\\
$1/g''$ & 0.258(8)& 0.4235 & 0.219(4)& 0.2415(83) & 0.249(8)\\
$\beta \mu^{({\rm exc})}_p$& & 3.161 & 2.8586(5) & 3.117(2) & 3.17(2) \\
\hline
\hline
\end{tabular}
\end{center}
\end{table}

\begin{table}[tbp!]
\caption{Estimates of several thermodynamic quantities 
for $q=1$, $\phi_c=0.1$, $\phi_p=0.8$. Definitions are given in 
App.~A of the paper. We report full-monomer (FM)
estimates and results obtained in CG models with 
$n=1$ and $n=4$ blobs.
We do not report the GFVT estimates, since this point belongs 
to the region in which GFVT predicts phase separation. For $\phi_c = 0.1$,
GFVT predicts 
the system to be homogeneous only up to $\phi_p\approx 0.72$.
}
\label{tabsuppl:t1_0p1_0p8}
\begin{center}
\begin{tabular}{cccc}
\hline\hline
& FM & $n=1$ & $n=4$ \\
\hline
$S_{pp,0}$ & 1.92(9) & 1.34(2)& 1.86(10)\\ 
$S_{cp,0}$ & $-$1.72(9)& $-$1.16(2)& $-$1.66(10)\\
$S_{cc,0}$ & 1.72(9)&  1.25(3)& 1.68(10)\\
$K_p$ & 6.45(8) & 5.08(3) &6.10(4)\\
$K_c$ & 18.7(2) & 14.19(8)& 17.7(2)\\
$\frac{\beta R_c^3}{\chi_T}$ & 1.68(2)& 1.309(7)& 1.59(1)\\
$1/g''$ & 0.28(1)& 0.196(4)& 0.27(2)\\
$\beta \mu^{({\rm exc})}_p$&  & 3.5692(5) & 3.978(4) \\
\hline
\hline
\end{tabular}
\end{center}
\end{table}

\begin{table}[tbp!]
\caption{Estimates of several thermodynamic quantities 
for $q=1$, $\phi_c=0.1$, $\phi_p=1.0$. Definitions are given in 
App.~A of the paper. We report full-monomer (FM)
estimates and results obtained in CG models with 
$n=1$ and $n=4$ blobs.
We do not report the GFVT estimates, since this point belongs 
to the region in which GFVT predicts phase separation. For $\phi_c = 0.1$,
GFVT predicts 
the system to be homogeneous only up to $\phi_p\approx 0.72$.
}
\label{tabsuppl:t1_0p1_1p0}
\begin{center}
\begin{tabular}{cccc}
\hline\hline
& FM & $n=1$ & $n=4$ \\
\hline
$S_{pp,0}$ & 3.08(25) & 1.36(4)& 2.97(25)\\ 
$S_{cp,0}$ & $-$2.95(25)& $-$1.25(4)& $-$2.82(25)\\
$S_{cc,0}$ & 2.98(25)&  1.36(4)& 2.85(25)\\
$K_p$ & 7.6(1) & 5.90(3) &7.27(4)\\
$K_c$ & 24.1(4) & 17.82(9)& 23.1(2)\\
$\frac{\beta R_c^3}{\chi_T}$ & 2.40(4)& 1.833(9)& 2.3(2)\\
$1/g''$ & 0.39(3)& 0.172(5)& 0.37(4)\\
$\beta \mu^{({\rm exc})}_p$& & 4.2796(5) & 4.862(5) \\
\hline \hline
\end{tabular}
\end{center}
\end{table}

\begin{table}[tbp!]
\caption{Estimates of several thermodynamic quantities 
for $q=1$, $\phi_c=0.2$, $\phi_p=0.2$. Definitions are given in 
App.~A of the paper. We report full-monomer (FM), 
GFVT estimates, and results obtained in CG models with 
$n=1$ and $n=4$ blobs.
}
\label{tabsuppl:t1_0p2_0p2}
\begin{center}
\begin{tabular}{ccccc}
\hline\hline
& FM & GFVT & $n=1$ & $n=4$ \\
\hline
$S_{pp,0}$ & 1.48(10) & 1.60& 1.44(1)& 1.54(2) \\ 
$S_{cp,0}$ & $-$0.66(5)& $-$0.76 &$-$0.619(8)& $-$0.691(1) \\
$S_{cc,0}$ & 0.44(2)& 0.53 & 0.428(5)& 0.461(8)\\
$K_p$ & 5.01(6) &  4.92 & 4.42(3) &4.95(3)\\
$K_c$ & 9.72(5) & 9.02  & 8.73(6)& 9.59(6)\\
$\frac{\beta R_c^3}{\chi_T}$ & 0.703(5)& 0.666 & 0.63(4)& 0.694(5)\\
$1/g''$ & 0.41(3)& 0.46 & 0.389(4)& 0.423(7)\\
$\beta \mu^{({\rm exc})}_p$& & 2.78 & 2.6064(6) & 2.763(3) \\
\hline \hline
\end{tabular}
\end{center}
\end{table}

\begin{table}[tbp!]
\caption{Estimates of several thermodynamic quantities 
for $q=1$, $\phi_c=0.2$, $\phi_p=0.4$. Definitions are given in 
App.~A of the paper. We report full-monomer (FM), 
GFVT estimates, and results obtained in CG models with 
$n=1$, $n=4$, and $n=10$ blobs.  Within GFVT, this state point 
is essentially on top of the fluid-fluid binodal and not far from the 
GFVT critical point $\phi_{c,\rm crit} = 0.178$, 
$\phi_{p,\rm crit} = 0.474$.
}
\label{tabsuppl:t1_0p2_0p4}
\begin{center}
\begin{tabular}{cccccc}
\hline\hline
& FM & GFVT & $n=1$ & $n=4$ & $n=10$\\
\hline
$S_{pp,0}$ & 2.68(7) & 12.2 & 2.00(3)& 2.62(5) & 2.63(3)\\ 
$S_{cp,0}$ & $-$1.49(4)& $-$7.56 &$-$1.07(2)& $-$1.46(30)& $-$1.48(2)\\
$S_{cc,0}$ & 0.95(2)& 4.80 & 0.70(1)& 0.92(2) & 0.94(1)\\
$K_p$ & 6.51(2) & 6.67 & 5.37(3) &6.45(3)& 6.63(2)\\
$K_c$ & 15.60(5) & 15.0 & 12.94(8)& 15.45(7) & 15.75(4)\\
$\frac{\beta R_c^3}{\chi_T}$ & 1.366(4)& 1.36 & 1.131(6)& 1.354(1) & 1.385(3)\\
$1/g''$ & 0.65(2)& 3.20 & 0.476(8)& 0.64(1) & 0.644(9)\\
$\beta \mu^{({\rm exc})}_p$& & 3.827 & 3.3926(7) & 3.746(3) & 3.77(1)\\
\hline \hline
\end{tabular}
\end{center}
\end{table}

\begin{table}[tbp!]
\caption{Estimates of several thermodynamic quantities 
for $q=1$, $\phi_c=0.3$, $\phi_p=0.2$. Definitions are given in 
App.~A of the paper. We report full-monomer (FM), 
GFVT estimates, and results obtained in CG models with 
$n=1$, $n=4$, and $n=10$ blobs.
}
\label{tabsuppl:t1_0p3_0p2}
\begin{center}
\begin{tabular}{cccccc}
\hline\hline
& FM & GFVT & $n=1$ & $n=4$  & $n=10$ \\
\hline
$S_{pp,0}$ & 2.37(9) &1.72 & 1.69(3)& 2.25(5) & 2.47(5)\\ 
$S_{cp,0}$ & $-$0.83(3)& $-$0.65 &$-$0.534(1)& $-$0.785(20) & $-$0.87(2)\\
$S_{cc,0}$ & 0.36(1)& 0.32 & 0.247(4)& 0.341(8)& 0.375(8)\\
$K_p$ & 8.60(3) & 8.50 & 6.88(5) &8.53(5) & 8.70(4)\\
$K_c$ & 19.00(7) &  17.57 & 16.2(1)& 18.9(1) & 19.19(8)\\
$\frac{\beta R_c^3}{\chi_T}$ & 1.772(6)& 1.625 & 1.49(1)& 1.76(1) & 1.790(8)\\
$1/g''$ & 0.57(2)& 0.404 & 0.392(8)& 0.541(1) & 0.60(2)\\
$\beta \mu^{({\rm exc})}_p$&  & 4.48 & 4.0829(8) & 4.529(5) & 4.58(2)\\
\hline
\hline
\end{tabular}
\end{center}
\end{table}

\begin{table}[tbp!]
\caption{Estimates of several thermodynamic quantities 
for $q=2$, $\phi_c=0.1$, $\phi_p=0.6$. Definitions are given in 
App.~A of the paper. We report full-monomer (FM), 
GFVT estimates, and results obtained in CG models with 
$n=1$, $n=4$, and $n=10$ blobs.
}
\label{tabsuppl:t2_0p1_0p6}
\begin{center}
\begin{tabular}{ccccccc}
\hline\hline
& FM & GFVT & $n=1$ & $n=4$ &$n=10$\\
\hline
$S_{pp,0}$ & 0.75(3) & 0.79 & 0.652(2)& 0.737(9) & 0.736(8)\\ 
$S_{cp,0}$ & $-$0.69(3)& $-$0.80  &$-$0.512(2)& $-$0.66(1) & $-$0.679(9) \\
$S_{cc,0}$ & 0.99(3)& 1.21 & 0.808(4)& 0.95(1) & 0.97(1)\\
$K_p$ & 6.81(6) & 6.86 & 5.29(7) &6.55(3) & 6.90(2)\\
$K_c$ & 5.13(3) & 4.77 & 4.14(8)& 5.00(3) & 5.20(2)\\
$\frac{\beta R_c^3}{\chi_T}$ & 0.245(2)& 0.237 & 0.1935(3)& 0.237(1)&0.248(1)\\ 
$1/g''$ & 0.38(2)& 0.432& 0.300(1)& 0.364(5) & 0.370(4)\\
$\beta \mu^{({\rm exc})}_p$& & 4.52 & 3.8541(3)& 4.402(2) & 4.46(3)\\
\hline
\hline
\end{tabular}
\end{center}
\end{table}

\begin{table}[tbp!]
\caption{Estimates of several thermodynamic quantities 
for $q=2$, $\phi_c=0.1$, $\phi_p=1.0$. Definitions are given in 
App.~A of the paper. We report full-monomer (FM), 
GFVT estimates, and results obtained in CG models with 
$n=1$, $n=4$, and $n=10$ blobs.  Within the GFVT
approximation, this point is not too far from the binodal (for $\phi_c = 0.1$
the system is homogeneeous up to $\phi_p\approx 1.3$) and the critical point
$\phi_{c,\rm crit} = 0.115$, $\phi_{p,\rm crit} = 1.205$.
}
\label{tabsuppl:t2_0p1_1p0}
\begin{center}
\begin{tabular}{cccccc}
\hline\hline
& FM & GFVT & $n=1$ & $n=4$ &$n=10$\\
\hline
$S_{pp,0}$ & 0.71(4) & 1.47 & 0.522(2)& 0.706(20) & 0.71(2)\\ 
$S_{cp,0}$ & $-$0.84(5)& $-$2.00 &$-$0.516(4)& $-$0.796(25) & $-$0.83(2)\\
$S_{cc,0}$ & 1.28(6)& 3.07 & 0.895(6)& 1.207(35) & 1.27(3)\\
$K_p$ & 9.8(1) & 9.75 & 6.74(1) &8.78(6) & 9.62(5)\\
$K_c$ & 7.94(7) & 7.43 & 5.46(1)& 7.30(6) & 7.88(4)\\
$\frac{\beta R_c^3}{\chi_T}$ & 0.481(5)& 0.469 & 0.332(6)& 0.436(3) & 0.476(2)\\
$1/g''$ & 0.458(25)& 1.07 & 0.307(2)& 0.44(1) & 0.460(1) \\
$\beta \mu^{({\rm exc})}_p$& &  6.52 & 5.2262(4)& 6.120(5) & 6.42(8)\\
\hline
\hline
\end{tabular}
\end{center}
\end{table}

\begin{table}[tbp!]
\caption{Estimates of several thermodynamic quantities 
for $q=2$, $\phi_c=0.2$, $\phi_p=0.4$. Definitions are given in 
App.~A of the paper. We report full-monomer (FM), 
GFVT estimates, and results obtained in CG models with 
$n=1$, $n=4$, and $n=10$ blobs.
}
\label{tabsuppl:t2_0p2_0p4}
\begin{center}
\begin{tabular}{cccccc}
\hline\hline
& FM & GFVT & $n=1$ & $n=4$ & $n=10$\\
\hline
$S_{pp,0}$ & 1.27(8) & 0.656 & 0.86(1)& 1.22(4) & 1.33(2)\\ 
$S_{cp,0}$ & $-$0.83(5)& $-$0.415 &$-$0.43(1)& $-$0.763(25) & $-$0.87(2)\\
$S_{cc,0}$ & 0.71(3)& 0.45 & 0.42(1)& 0.65(2) & 0.74(1)\\
$K_p$ & 11.35(7) & 10.56 & 7.36(4) &10.5(1) & 11.30(7)\\
$K_c$ & 8.07(4) & 7.14 & 6.20(7)& 7.77(8) & 8.05(5)\\
$\frac{\beta R_c^3}{\chi_T}$ & 0.521(3)& 0.467 & 0.383(3)& 0.495(5)& 0.520(3) \\
$1/g''$ & 0.29(2)& 0.151 & 0.179(3)& 0.274(9) & 0.306(3)\\
$\beta \mu^{({\rm exc})}_p$& & 6.58 & 5.519(1)& 6.713(5) & 6.97(3)\\
\hline
\hline
\end{tabular}
\end{center}
\end{table}

\begin{table}[tbp!]
\caption{Estimates of several thermodynamic quantities 
for $q=2$, $\phi_c=0.2$, $\phi_p=0.8$. Definitions are given in 
App.~A of the paper. We report full-monomer (FM), 
GFVT estimates, and results obtained in CG models with 
$n=1$, $n=4$, and $n=10$ blobs.
}
\label{tabsuppl:t2_0p2_0p8}
\begin{center}
\begin{tabular}{cccccc}
\hline\hline
& FM & GFVT & $n=1$ & $n=4$ & $n=10$\\
\hline
$S_{pp,0}$ & 2.6(3) & 1.43 & 0.701(8)& 1.55(9) & 2.6(1)\\ 
$S_{cp,0}$ & $-$2.2(3)& $-$1.33 &$-$0.472(8)& $-$1.25(8) & $-$2.2(1)\\
$S_{cc,0}$ & 2.0(2)& 1.39 & 0.510(9)& 1.15(7) & 2.0(1)\\
$K_p$ & 15.2(2) & 15.27 & 8.79(2) &13.2(2) & 14.8(1)\\
$K_c$ & 12.2(1) & 11.05 & 7.72(4)& 11.0(2) & 11.9(1)\\
$\frac{\beta R_c^3}{\chi_T}$ & 0.946(9)&0.892  & 0.579(2)& 0.84(1) & 1.00(5)\\
$1/g''$ & 1.0(1)& 0.593 & 0.241(4)& 0.58(4) & 1.00(5)\\
$\beta \mu^{({\rm exc})}_p$& & 9.29 & 6.920(7)& 8.81(1)  & 9.42(5)\\
\hline
\hline
\end{tabular}
\end{center}
\end{table}

\clearpage

\begin{table}[tbp!]
\caption{Estimates of several thermodynamic quantities 
for $q=2$, $\phi_c=0.3$, $\phi_p=0.2$. Definitions are given in 
App.~A of the paper. We report full-monomer (FM), 
GFVT estimates, and results obtained in CG models with 
$n=1$, $n=4$, and $n=10$ blobs.
}
\label{tabsuppl:t2_0p3_0p2}
\begin{center}
\begin{tabular}{cccccc}
\hline\hline
& FM & GFVT & $n=1$ & $n=4$ & $n=10$ \\
\hline
$S_{pp,0}$ & 1.5(1) & 0.48 & 0.863(4)& 1.38(4) & 1.58(3)\\ 
$S_{cp,0}$ & $-$0.60(4)& $-$0.147 &$-$0.197(1)& $-$0.50(2) & $-$0.62(1)\\
$S_{cc,0}$ & 0.319(15)& 0.136 & 0.143(2)& 0.263(7) & 0.326(5)\\
$K_p$ & 18.74(7) & 15.01 & 9.8(1) &16.6(2) & 18.6(1) \\
$K_c$ & 13.29(4) & 12.07  & 10.9(2)& 12.8(1) & 13.22(5)\\
$\frac{\beta R_c^3}{\chi_T}$ & 1.063(3)& 0.954 & 0.84(1)& 1.015(10) & 1.057(4) \\
$1/g''$ & 0.124(8)& 0.037 & 0.0648(4)& 0.111(4) & 0.128(2)\\
$\beta \mu^{({\rm exc})}_p$& & 8.58 & 7.4151(7)& 9.778(8)& 10.27(2)\\
\hline
\hline
\end{tabular}
\end{center}
\end{table}


\begin{thebibliography}{99}
\bibitem{footnote-w}
Simulations have been performed at $w = 0.505838$, as in our previous
works. This value is optimal within errors.
The optimal value $w^*$ at which the leading scaling corrections
are absent was estimated in Ref.~\onlinecite{CMP-06-suppl}, obtaining
$w^* = 0.48\pm 0.02$.
\end{thebibliography}
\end{document}